\documentclass[a4paper,twoside,english,aps,preprint]{revtex4}
\usepackage[T1]{fontenc}
\usepackage[latin1]{inputenc}
\usepackage{subfigure}
\usepackage{amsmath}
\usepackage{graphicx}
\usepackage{amssymb}

\makeatletter

\newcommand{\noun}[1]{\textsc{#1}}
\providecommand{\tabularnewline}{\\}

\newcommand{\B}{\mathbf{B}}

\newcommand{\V}{\mathcal{V}}

\newcommand{\pj}{\boldsymbol {\pi}}
\newcommand{\xib}{\boldsymbol {\xi}}
\newcommand{\etab}{\boldsymbol {\eta}}
\newcommand{\zetab}{\boldsymbol {\zeta}}
\newcommand{\al} { \tilde{\alpha} }
\newcommand{\Sym} { \mathrm{S} }
\newcommand{\ca} {\mathcal{A}}
\newcommand{\omv} {\langle\omega\rangle_v }
\newcommand{\av} {\langle A \rangle_v }

\usepackage{babel}
\makeatother
\begin{document}

\title{$N$-Electron Giant Dipole States in Crossed Electric and Magnetic
Fields}

\author{Sascha Zöllner}

\email{sascha@pci.uni-heidelberg.de}

\thanks{}

\affiliation{Theoretische Chemie, Institut f\"{u}r Physikalische Chemie, Universit\"{a}t
Heidelberg, INF 229, 69120 Heidelberg, Germany }

\author{Hans-Dieter Meyer}

\email{dieter@pci.uni-heidelberg.de}

\affiliation{Theoretische Chemie, Institut f\"{u}r Physikalische Chemie, Universit\"{a}t
Heidelberg, INF 229, 69120 Heidelberg, Germany}

\author{Peter Schmelcher}

\email{peter@pci.uni-heidelberg.de}

\affiliation{Theoretische Chemie, Institut f\"{u}r Physikalische Chemie, Universit\"{a}t
Heidelberg, INF 229, 69120 Heidelberg, Germany}

\affiliation{Physikalisches Institut, Universit\"{a}t Heidelberg, Philosophenweg
12, 69120 Heidelberg, Germany}

\begin{abstract}
Multi-electron giant dipole resonances of atoms in crossed electric
and magnetic fields are investigated. Stationary configurations corresponding
to a highly symmetric arrangement of the electrons on a decentered
circle are derived, and a normal-mode and stability analysis are performed.
A classification of the various modes, which are dominated by the
magnetic field or the Coulomb interactions, is provided. Based on
the \noun{mctdh} approach, we carry out a six-dimensional wave-packet
dynamical study for the two-electron resonances, yielding in particular
lifetimes of more than $0.1\mu$s for strong electric fields. 
\end{abstract}
\maketitle

\section{introduction}

Interactions of atoms with strong electric and magnetic fields, in
particular \emph{crossed} fields, have long been the focal point of
various research efforts \cite{burkova76,rau86,melezhik93,raithel95,zhang95,floethmann96,haggerty97}.
The conflicting symmetries inherent in the electronic motion in atomic
and external fields make for a rich variety of phenomena. In this
light they are not only interesting from a theoretical standpoint,
such as the hydrogen atom, which constitutes a paradigm for a chaotic
system \cite{Friedrich97,Herold81:751,Friedrich89:37,Schmelcher98,Gallagher94}:
Their study also had a major impact on a number of other fields, ranging
from semiconductor physics to astrophysics. 

However, for a long time theoretical investigations did not go beyond
the stage of treating the nucleus as {}`frozen', the so-called \emph{infinite-mass
approximation}. The model underlying this picture is the concept that
the atomic motion could be \emph{separated} into that of the center
of mass and the one relative to it. While this is perfectly justified
for translationally invariant systems---going along with the conservation
of the total momentum---it ceases to be true in the presence of magnetic
fields. Nonetheless, a weaker version of this procedure can be established
in the case of (neutral) atoms in homogeneous fields \cite{Lamb59,Avr78:431,Herold81:751,Hirsch83}.
In this \emph{pseudo-separation}, a conserved quantity termed \emph{pseudo-momentum}---the
total momentum plus a field-dependent compensation---allows to obtain
a Hamiltonian that depends on the relative coordinates only. However,
that relative motion strongly depends on the center of mass via the
pseudo-momentum. This coupling gives rise to some new and interesting
effects, such as \emph{chaotic diffusion} of the center of mass \cite{Schm92:305,Schm92:2697},
or \emph{dynamical self-ionization} \cite{Schm95:662,melezhik00}
for the case of ions.

Yet one of the most prominent among these \emph{finite-mass} effects
is the existence of \emph{giant dipole states}, where the electrons
can be decentered from the nucleus by many $10,000$ atomic units
\cite{Baye92,Dzyaloshinskii92}. The systematic base for this was
the gauge-independent generalized potential first derived for hydrogen
by Dippel \emph{et~al.} \cite{Dip94:4415} and extended to arbitrary
atoms by Schmelcher \cite{Schm01}. It has been applied extensively
to study giant dipole states in the two-body case; e.g., hydrogen
\cite{Dip94:4415} and positronium \cite{Ack98:1129,Ack97}, where
the large inter-particle distance prevents annihilation for up to
several years. By contrast, the extension to the \emph{multi-electron}
case has attracted only limited attention. There have been indications
for two-electron quasistable giant-dipole states based on a local
semiclassical analysis \cite{Schm01}. The aim of this paper is to
both provide a local analysis of general $N$-electron system and
investigate the stability of the \emph{full} quantum-mechanical system.
In order to first analyze the $N$-electrons case analytically, we
look for stationary points of the so-called \emph{generalized potential}
of the effective relative motion (the {}`giant-dipole configuration')
so as to carry out a normal-mode analysis about these points. Equipped
with this insight into the \emph{local} behavior, we study the \emph{exact}
system $N=2$ numerically using wave-packet propagation. Here we resort
to the \noun{mctdh} method (Multi-Configuration Time-Dependent Hartree)
\cite{mey03:251,mey98:3011,bec00:1}.

This article is organized as follows. In Sec.~\ref{sec:The-N-electron-problem},
the theoretical framework of the pseudo-separation is reviewed and
applied to derive the \emph{generalized potential} of the relative
motion. Sec.~\ref{sec:Stationary-configurations} then deals with
the stationary points of the generalized potential as a base for the
analysis of giant dipole resonances. In the subsequent section, the
normal-mode analysis for $N$ electrons is carried out. To this end,
the local equations of motion are derived (\ref{sec:Equations-of-motion}),
whose eigenmodes and -vectors are computed numerically (\ref{sub:Results-and-discussion})
as the solutions of a quadratic eigenvalue problem. Sec.~\ref{sec:Wave-packet-dynamical-study}
contains a wave-packet dynamical study of the two-electron system
based on the \noun{mctdh} method. Results on the stability, spectral
properties and lower bounds for the lifetimes of the resonances are
presented.

\section{The $N$-electron atom in crossed electric and magnetic fields\label{sec:The-N-electron-problem}}

\newcommand{\exv}[1]{\langle#1\rangle}
The Hamiltonian of an atom, consisting of $N$ electrons (of mass
$m$) and a nucleus (mass $M_{0}$) interacting via the Coulomb potential
$V$, reads in the laboratory frame:\[
H=\frac{1}{2m}\sum_{i=1}^{N}\left(\mathbf{p}_{i}-e\mathbf{A}(\mathbf{r}_{i}^{\prime})\right)^{2}+\frac{1}{2M_{0}}\left(\mathbf{p}_{0}+Ze\mathbf{A}(\mathbf{r}_{0}^{\prime})\right)^{2}-e\sum_{i=1}^{N}\mathbf{E}\cdot\mathbf{r}_{i}^{\prime}+Ze\mathbf{E}\cdot\mathbf{r}_{0}^{\prime}+V.\]
Here an electron (index $i\le N$) has the electric charge $e=-|e|$;
the atomic number $Z$ equals the number of electrons $N$ in the
neutral case treated here. We consider static electric and magnetic
fields, which are accounted for by the \emph{electro-static potential}
$\phi(\mathbf{r})=-\mathbf{E}\cdot\mathbf{r}$ and the \emph{vector
potential} $\mathbf{A}(\mathbf{r})$, whose gauge is not fixed here.

In the field-free case, the translation invariance of the system---going
along with the conservation of the total momentum---guarantees a complete
separation of the center-of-mass and the relative motion. The vector
potential now breaks that symmetry. For the special case of homogeneous
fields, though, it was shown that a so-called \emph{pseudo-separation}
of the center-of-mass motion is possible . The key is that even though
the total momentum is not conserved, the total \emph{pseudo-momentum}
$\mathbf{K}:=\sum_{a=0}^{N}\mathbf{k}_{a}$ defined in terms of the
\emph{one-particle pseudo-momenta}\[
\mathbf{k}_{i}:=m\mathbf{v}_{i}-e\mathbf{r}_{i}^{\prime}\times\mathbf{B}\]
---and likewise for the nucleus---is constant by construction \cite{Avr78:431,Hirsch83,Herold81:751}.
It can be thought of as the \emph{kinetic} center-of-mass (CM) momentum
plus a magnetic-field-dependent compensation, which trivially reduces
to the total momentum in the case $B=0$. Moreover, its components
$\{ K_{\alpha}\}$ mutually commute for a neutral system, making them
commensurable constants of motion. This is crucial in performing a
gauge-independent pseudo-separation of the CM motion \cite{Dip94:4415,Schm01},
which consists of the following steps:

\begin{enumerate}
\item Transformation to CM and coordinates relative to the nucleus, $\{\mathbf{R}_{CM},\mathbf{r}_{i}\}$,
yielding $H_{CM}$ and $\mathbf{K}_{CM}$.
\item Choose the total wave function to be a simultaneous eigenstate of
the pseudomomentum with eigenvalue $\mathbf{K}$: $(\mathbf{K}_{CM}-\mathbf{K})\Psi_{\mathbf{K}}=0$.
The result is of the form \begin{equation}
\Psi_{\mathbf{K}}(\mathbf{R}_{CM},\{\mathbf{r}_{j}\})=\mathcal{U}(\mathbf{R}_{CM},\{\mathbf{r}_{j}\})\phi(\{\mathbf{r}_{j}\}),\label{eq:pseudo_psi}\end{equation}
where $\phi$ is an arbitrary function of the relative coordinates,
and $\mathcal{U}$ denotes a unitary operator.
\item Finally reduce the full Schrödinger equation to an effective one for
the relative coordinates via\begin{equation}
(\mathcal{H}-E)\phi=0,\mathrm{\quad with\quad}\mathcal{UHU}^{-1}=H_{CM}.\label{eq:pseudo_H}\end{equation}

\end{enumerate}
It has to be emphasized that gauge independence is an essential quality
of the pseudo-separation scheme outlined here. A major drawback inherent
in any fixed-gauge approach is that one cannot identify gauge-independent
terms in the effective Hamiltonian (\ref{eq:pseudo_H}), making the
analysis less systematic.

In order to derive this effective Hamiltonian in a gauge-invariant
way, we decompose the vector potential according to \[
\mathbf{A}(\mathbf{r})=\frac{1}{2}\mathbf{B\times r}+\nabla\Lambda(\mathbf{r}).\]
The transform $\mathcal{U}$ and hence the effective Hamiltonian $\mathcal{H}=\mathbb{\mathcal{T}}+\mathcal{V}$
is then determined only up to some $\Lambda$-dependent function $f(\{\mathbf{r}_{j}\})$
of the relative coordinates which vanishes by construction for $\nabla\Lambda=0$
\cite{Schm01}: \begin{eqnarray}
\mathcal{T} & = & \frac{1}{2m}\sum_{i=1}^{N}\left(\mathbf{p}_{i}-\frac{e}{2}\mathbf{B}\times\left[\mathbf{r}_{i}-2\frac{m}{M}\sum_{j=1}^{N}\mathbf{r}_{j}\right]+e\nabla_{i}f\right)^{2}+\nonumber \\
 &  & \frac{1}{2M_{0}}\left(\sum_{j=1}^{N}\left[\mathbf{p}_{j}-\frac{q}{2N}\mathbf{B}\times\mathbf{r}_{j}+e\nabla_{j}f\right]\right)^{2}\label{eq:general_kin}\\
\mathcal{V} & = & \frac{1}{2M}\left(\mathbf{K}-e\mathbf{B}\times\sum_{j=1}^{N}\mathbf{r}_{j}\right)^{2}-e\sum_{i=1}^{N}\mathbf{E}\cdot\mathbf{r}_{i}+V,\label{eq:general_pot}\end{eqnarray}
with $q=Ne(M_{0}-Nm)/M$. 

$\mathcal{T}$ represents the (gauge-dependent) \emph{kinetic energy}
of the relative motion; it depends on the gauge via $f$. As opposed
to that, the \emph{generalized potential} $\mathcal{V}$ is manifestly
gauge-independent and is therefore identified as an effective potential
for the electronic motion at a given value of the total pseudo-momentum.
Beyond the usual Coulomb potential and the \emph{Stark terms} due
to the external electric field, we have an expression corresponding
to the CM-kinetic energy,\begin{equation}
\frac{1}{2M}\left(\mathbf{K}-e\mathbf{B}\times\sum_{j=1}^{N}\mathbf{r}_{j}\right)^{2}=\frac{\mathbf{K}^{2}}{2M}+\frac{e^{2}}{2M}\left(\mathbf{B}\times\sum_{j=1}^{N}\mathbf{r}_{j}\right)^{2}-e\left(\frac{\mathbf{K\times B}}{M}\right)\cdot\sum_{j=1}^{N}\mathbf{r}_{j}.\label{eq:CM-Ekin}\end{equation}
 The first term, the analogue of the field-free CM energy, is merely
a constant. The second expression is a diamagnetic term, which acts
as a confining harmonic potential on the electrons' center of mass,
$\frac{1}{N}\sum_{j=1}^{N}\mathbf{r}_{j}$, with a frequency $\boldsymbol{\omega}{}_{CM}\equiv Ne\mathbf{B}/M$
perpendicular to the magnetic field. Most notably, the center of mass---unlike
in the field-free case---actually couples to the electrons (i.e.,
to their center of mass) via the \emph{motional electric field} $\mathbf{E}_{M}:=\frac{\mathbf{K\times B}}{M}$.
For crossed fields, it is therefore inviting to combine the Stark
terms arising from both the external and the motional electric field
and account for them by defining the \emph{effective pseudomomentum}\[
\mathbf{K}':=\mathbf{K}+M\frac{\mathbf{E\times B}}{B^{2}}\equiv\mathbf{K}-M\mathbf{u},\]
$\mathbf{u}$ being the classical drift velocity. The Hamiltonian
then reads \cite{Dip94:4415,Ack98:1129}\[
\mathcal{H}(\mathbf{K},\mathbf{E})=\mathcal{H}(\mathbf{K}',\mathbf{0})+\mathbf{K\cdot u}-\frac{1}{2}M\mathbf{u}^{2}.\]
We have thus reduced our problem of crossed fields to a purely magnetic
one. In what follows, we will always deal with $\mathcal{H}(\mathbf{K}',\mathbf{0})$
for simplicity.

The generalized potential derived above gives rise to many effects,
among the most prominent ones being the existence of \emph{giant dipole
states} (GDS).  The concern of the following two sections is to derive
stationary points of that potential which accomodate these states,
and to classify their normal modes.

\section{Stationary configurations of the generalized potential\label{sec:Stationary-configurations}}

With the gauge-independent generalized potential at hand, we can now
start our search for decentered stationary points, which we will identify
as candidates for giant dipole resonances. To facilitate the search
for stationarities, we will first choose convenient coordinates (Subsection
\ref{sec:The-electronic-center}) which present a suitable extension
of the case $N=2$ \cite{Schm01}. We will then derive and solve the
stationarity conditions by an educated guess in \ref{sec:Stationary-Points}.

\subsection{The electronic center of mass frame\label{sec:The-electronic-center}}

As we know from the studies on the one-electron case \cite{Schm01,Ack98:1129,Dip94:4415},
it is the CM-kinetic energy (\ref{eq:CM-Ekin}) that is responsible
for the decentered states. It was pointed out in Eq. (\ref{eq:general_pot})
already that this term depends on the combination $\mathbf{R}=\frac{1}{N}\sum_{j=1}^{N}\mathbf{r}_{j},$
which we refer to as the \emph{electronic center of mass} \emph{(ECM}).
It is thus natural to introduce the ECM as a new coordinate. In addition,
we will choose $N-1$ vectors relative to the ECM, i.e. to decompose
$\mathbf{r}_{i}=\mathbf{R}+\mathbf{s}_{i}$, where the $\{\mathbf{s}_{i}\}$
only account for all but one (say, the $N$th) electron's relative
position:\begin{eqnarray*}
(\mathbf{r}_{1},\dots,\mathbf{r}_{N}) & \mapsto & (\mathbf{R};\mathbf{s}_{1},\dots,\mathbf{s}_{N-1})\\
\mathbf{R} & := & \frac{1}{N}\sum_{i=1}^{N}\mathbf{r}_{i}\\
\mathbf{s}_{i} & := & \mathbf{r}_{i}-\frac{1}{N}\sum_{j=1}^{N}\mathbf{r}_{j}\qquad(i=1,\dots,N-1),\end{eqnarray*}
Introducing the conjugate momenta $\mathbf{P}=-i\frac{\partial}{\partial\mathbf{R}},\,\boldsymbol{\pi}_{j}=-i\frac{\partial}{\partial\mathbf{s}_{j}}$,
the Hamiltonian $\mathcal{H}=\mathcal{T}+\mathcal{V}$ becomes (with
$f=0$ for simplicity)\begin{eqnarray}
\mathcal{T} & = & \frac{1}{2\mu_{N}}(\mathbf{P}-\frac{q}{2}\B\times\mathbf{R})^{2}+\frac{1}{2m}\sum_{i<N}\left(\pj_{i}-\frac{1}{N}\sum_{k<N}\pj_{k}-\frac{e}{2}\B\times\mathbf{s}_{i}\right)^{2}+\nonumber \\
 &  & \frac{1}{2m}\left(\frac{1}{N}\sum_{i<N}\pj_{i}-\frac{e}{2}\B\times\sum_{i<N}\mathbf{s}_{i}\right)^{2}\label{eq:kinetic_ECM}\\
\mathcal{V} & = & \frac{1}{2M}(\mathbf{K'}-e\B\times N\mathbf{R})^{2}-Ze^{2}\sum_{a=1}^{N}\frac{1}{r_{a}}+e^{2}\sum_{a<b\le N}\frac{1}{s_{ab}},\label{eq:potential_ECM}\end{eqnarray}
where we introduced the shorthands\begin{eqnarray*}
\mathbf{s}_{N} & \equiv & \mathbf{r}_{N}-\mathbf{R}=-\sum_{i=1}^{N-1}\mathbf{s}_{i}\\
r_{a} & \equiv & |\mathbf{r}_{a}|=|\mathbf{R}+\mathbf{s}_{a}|\\
s_{ab} & \equiv & |\mathbf{s}_{a}-\mathbf{s}_{b}|\quad(a<b\le N)\end{eqnarray*}
as well as the reduced mass $\mu_{N}^{-1}=(Nm)^{-1}+M_{0}^{-1}$.
It is reassuring that the kinetic energy can be shown to have the
intuitive form\[
\mathcal{T}=\frac{1}{2}\mu_{N}\mathbf{\dot{R}}^{2}+\frac{1}{2}\sum_{a=1}^{N}m\dot{\mathbf{s}}_{a}^{2}.\]

\subsection{Existence of stationary points\label{sec:Stationary-Points}}

Given the advantageous form of the Hamiltonian (\ref{eq:kinetic_ECM}-\ref{eq:potential_ECM}),
we are now ready to look for stationary points of the potential. If
we define $Q=(\mathbf{R},\mathbf{s}_{1},\dots,\,\mathbf{s}_{N-1})^{T}\in\mathbb{R}^{3N}$,
then we seek some $Q_{0}$ such that\[
\frac{\partial\mathcal{V}}{\partial Q}(Q_{0})=0.\]
This \emph{stationarity condition} leads to the rather intricate set
of equations, \begin{eqnarray}
\mathbf{0}=\frac{\partial\mathcal{V}}{\partial\mathbf{R}}(Q) & = & \frac{Ne\mathbf{B}}{M}\times(\mathbf{K'}-Ne\mathbf{B\times R})+\label{eq:DV_R}\\
 &  & Ze^{2}\left[\mathbf{R}\left(\sum_{i<N}\frac{1}{r_{i}^{3}}+\frac{1}{r_{N}^{3}}\right)+\sum_{i<N}\mathbf{s}_{i}\left(\frac{1}{r_{i}^{3}}-\frac{1}{r_{N}^{3}}\right)\right]\nonumber \\
\mathbf{0}=\frac{\partial\mathcal{V}}{\partial\mathbf{s}_{i}}(Q) & = & Ze^{2}\left[\mathbf{R}\left(\frac{1}{r_{i}^{3}}-\frac{1}{r_{N}^{3}}\right)+\frac{\mathbf{s}_{i}}{r_{i}^{3}}-\frac{\mathbf{s}_{N}}{r_{N}^{3}}\right]-\label{eq:DV_i}\\
 &  & e^{2}\left[\mathbf{s}_{i}\left(\sum_{{{k<N\atop k\neq i}}}\frac{1}{s_{ik}^{3}}+\frac{1}{s_{iN}^{3}}\right)-\sum_{{{k<N\atop k\neq i}}}\frac{\mathbf{s}_{k}}{s_{ik}^{3}}+\sum_{k<N}\mathbf{s}_{k}\left(\frac{1}{s_{iN}^{3}}+\frac{1}{s_{kN}^{3}}+\sum_{j<N}\frac{1}{s_{jN}^{3}}\right)\right].\nonumber \end{eqnarray}
(Throughout this subsection, the subscript in $Q_{0}$ is omitted
for mnemotechnical reasons.)

Equations (\ref{eq:DV_R}) and (\ref{eq:DV_i}) constitute a $3N$-dimensional
non-linear system of equations that has little hope of being solved
analytically in complete generality. However, one can try to find
\emph{special solutions} by an educated guess. For instance, one immediately
sees that Eq. (\ref{eq:DV_R}) simplifies considerably if $r_{i}=r_{N}$
for all $i$. At that stage, all electrons are distributed on the
shell of a sphere centered around the nucleus. It is tempting to go
one step further and require $\mathbf{R}\perp\mathbf{s}_{i}$ so as
to {}`decouple' the two types of coordinates. This motivates our
set of assumptions I:\begin{eqnarray}
r_{i}=r_{N} & =: & r,\:\mathbf{R}\cdot\mathbf{s}_{i}=0\quad\forall i<N.\label{eq:ass}\end{eqnarray}
 Before summarizing the immediate consequences, let us specify a basis
$\{\xib,\etab,\zetab\}$ such that the magnetic field points along
the $\zetab$ direction:\begin{eqnarray*}
\zetab & = & \mathbf{B}/B\\
\etab & = & \mathbf{K}'/K'\\
\xib & = & \etab\times\zetab/|\etab\times\zetab|.\end{eqnarray*}
(For convenience, we assume $\angle(\mathbf{K}',\mathbf{B})=90°$
in addition to the premise of crossed fields. This may be achieved
by preparing a state with $K\simeq0$ so that only the external electric
field is relevant.) In this basis, write the extremal ECM vector in
terms of $\mathbf{R}\cdot\xib=X,\,\mathbf{R}\cdot\etab=Y,\,\mathbf{R}\cdot\zetab=Z$.

If (\ref{eq:ass}) holds, then the extremal condition (\ref{eq:DV_R}),
$\frac{\partial\mathcal{V}}{\partial\mathbf{R}}\big|_{Q}=\mathbf{0}$,
is equivalent to

\begin{enumerate}
\item $Y=Z=0$, that is $\mathbf{R}=(X,0,0)^{T},$ where
\item $X$ satisfies $P_{r}(X)=-\frac{NeK'B}{M}+\frac{(NeB)^{2}}{M}X+NZe^{2}\frac{X}{r^{3}}=0$.
\item Moreover, there is a set of angles $\phi_{i}$ such that $\mathbf{s}_{i}=s\,(0,\cos\phi_{i},\sin\phi_{i})^{T}$
$\forall i<N$.
\end{enumerate}
By (1.), if there is a solution, then the electronic center of mass
will be aligned along the $\xib$ axis, that is the decentering takes
place anti-parallel to the electric field. Based on this, statement
(2.) gives the (implicit) condition for the existence of solutions
$X$. Finally, (3.) reflects a constraint on all relative coordinates
$\mathbf{s}_{a}$: they have the same norm $s$ and---due to their
orthogonality to $\mathbf{R}$---can be placed arbitrarily on a circle
about $\mathbf{R}$. For symmetry reasons, we demand that all relative
coordinates be distributed uniformly over the circle, which amounts
to the \textbf{}assumption \textbf{}II:\begin{equation}
\phi_{k}=\Phi_{N}k+\Delta\;\forall k,\quad\Phi_{N}\equiv\frac{2\pi}{N}.\label{eq:ass-2}\end{equation}
This procedure determines the circular configuration only up to a
global rotation by an angle $\Delta\in[0,2\pi).$

The above guess allows us to fix the ratio of $|X|$ and $s$, or
$r$ and $s$, respectively, so as to meet Eq. (\ref{eq:DV_i}):\begin{eqnarray}
r & = & s\sqrt[3]{\frac{4N}{\sum_{k<N}(\sin\frac{\Phi k}{2})^{-1}}}=:\alpha_{N}s\label{eq:alpha}\\
r & = & (1-\alpha_{N}^{-2})^{-\frac{1}{2}}|X|=:\tilde{\alpha}_{N}|X|.\label{eq:alpha1}\end{eqnarray}
Pictorially, $\alpha_{N}$ is related to the angle $\delta_{N}$ between
$\mathbf{R}$ and $\mathbf{r}_{i}$ via $\sin\delta_{N}=\alpha_{N}^{-1}$.
To be explicit, we have computed the parameters for some relevant
$N$ below:

\begin{center}\begin{tabular}{|c|c|c|}
\hline 
$N$&
$\alpha_{N}=\frac{r}{s}$&
$\tilde{\alpha}_{N}=\frac{r}{|X|}$\tabularnewline
\hline
\hline 
2&
2&
$\frac{2}{\sqrt{3}}=$1.15470054...\tabularnewline
\hline 
3&
$\sqrt{3}=1.7320508...$&
$\sqrt{\frac{2}{3}}=1.22474487...$\tabularnewline
\hline 
4&
1.6107688...&
1.27558847...\tabularnewline
\hline 
5&
1.5372471...&
1.31666491...\tabularnewline
\hline
\end{tabular}\\
\end{center}

The connection between $r$ and $X$ now also enables us to turn the
implicit equation $P_{r}(X)=0$ above into a cubic equation\begin{equation}
X^{3}+X^{2}\frac{K'}{NB}-\frac{M}{\al^{3}B^{2}}=0,\quad X<0.\label{eq:P(X)}\end{equation}
Eq. (\ref{eq:P(X)}) has two distinct solutions $X_{1/2}(K;B)$ provided
that $K'>K_{cr}\equiv\frac{3N}{\al}\sqrt[3]{\frac{MB}{4}}$:\begin{eqnarray}
X_{\lambda} & = & \frac{K'}{3NB}\Big(2\cos\left(\frac{\theta+\lambda\cdot2\pi}{3}\right)-1\Big),\qquad\lambda=1,\,2\label{eq:X_lambda}\\
\cos\theta & \equiv & 2\left(\frac{K_{cr}}{K'}\right)^{3}-1.\nonumber \end{eqnarray}

In other words: for any $B$ there is a critical value $K_{cr}$ for
$K'$ below which $\frac{\partial\mathcal{V}}{\partial X}(X\boldsymbol{\xi},\mathbf{s}_{i}^{(0)})=0$
has complex solutions. In an experiment, one therefore has to control
the value of $K'=|\mathbf{K}+M\mathbf{E\times B}/B^{2}|$. Averbukh
et al. \cite{Ave99:3695} suggested how this could be achieved experimentally.
The underlying idea is to prepare an atom with low $K$ (requiring
a slow CM motion), and use the electric field $E$ to control the
critical parameter. This way an initially prepared Coulomb-Rydberg
state is transformed into a decentered state localized in the extremum.

\begin{figure}
\begin{center}\includegraphics[%
  width=10cm,
  keepaspectratio]{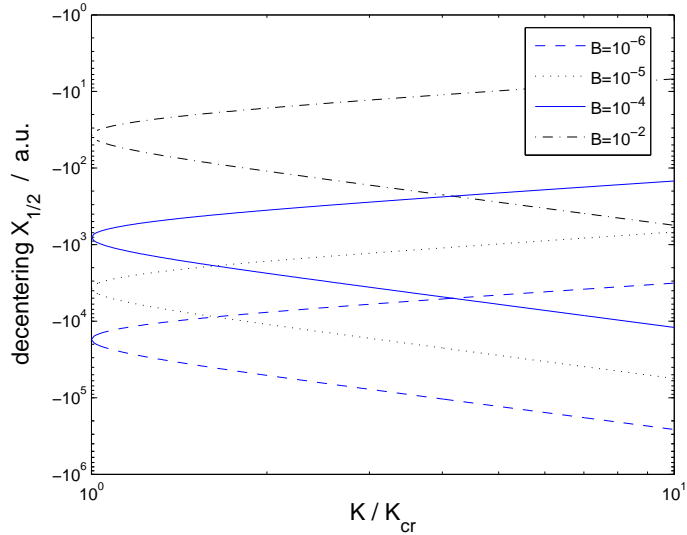}\end{center}
\vspace{-0.5cm}

\caption{\label{cap:decenter}The extremal component $X_{1/2}$ as a function
of $K/K_{cr}$ for different field strengths $B/\mathrm{a.u.}$ The
critical point $X(K_{cr})$ corresponds to the vertex where $X_{1}$
(lower graph) and $X_{2}$ (upper graph) meet.}
\end{figure}

To gain a better understanding of the two solutions, their behavior
is illustrated in Figure \ref{cap:decenter}. For fixed $B$, both
lines meet at the vertex \emph{}$X_{c}\equiv X_{1}(K_{cr})=X_{2}(K_{cr})=-K_{cr}/NB$.
From there on, the lower branch $X_{1}(K)$ tends to $-\infty$ monotonically
(which we identify with a proper decentering), whereas the upper curve
$X_{2}(K)$ goes to zero (i.e., the decentering gets lost). %
For stronger magnetic fields, in turn, the {}`decentering region'
around $X_{c}$ shrinks. We remark that the decentering is virtually
insensitive to the number of electrons, $N$.

\begin{figure}
\begin{center}\vspace{-0.5cm}\includegraphics[%
  width=9cm]{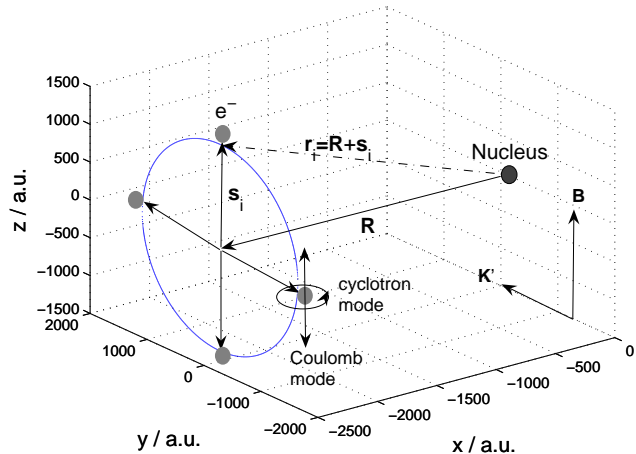}\vspace{-0.5cm}\end{center}

\caption{A giant-dipole configuration for $N=4$ electrons ($B=10^{-4},\, K=2K_{cr}$).
The ECM $\mathbf{R}=(X,0,0)$ is decentered, with the relative vectors
$\mathbf{s}_{i}$ confined to a circle as indicated. Also shown is
the electronic vector relative to the nucleus $\mathbf{r}_{i}=\mathbf{R}+\mathbf{s}_{i}$.
A typical example of the cyclotron and the motion due to a Coulomb
mode is indicated (see Subsec. \ref{sub:General-Classification}).
\label{cap:giant-dipole-configuration}}
\end{figure}
Let us now wrap up our results. Every stationary electronic vector
possesses the orthogonal decomposition\[
\mathbf{r}_{i}=X\xib+s(0,\cos\phi_{i},\sin\phi_{i})^{T}.\]
 This means that all electrons have a common position along $\xib$---which
we refer to as the \emph{decentering}---while they are confined to
a highly symmetric circular configuration perpendicular to $\xib$.
This circular configuration in turn is determined only up to an overall
rotation $\Delta$. A generic setup is shown in Figure \ref{cap:giant-dipole-configuration}.

The extremal solution exists only if the effective pseudomomentum
$K'$ (controlled via the electric field) exceeds some critical value.
The more it does so, the more distinct the decentering of the ECM
will be.

\section{Normal-mode analysis of $N$-electron giant dipole states\label{sec:Normal-mode-analysis}}

As argued in the previous section, we expect the stationary configurations
to be candidates for giant dipole resonances \emph{}with certain lifetimes\emph{.}
Preceding a numerical study, we first seek to obtain some insight
into the \emph{local} stability of the system, i.e., their behavior
in the vicinity of the extrema. To that end, we will carry out a normal-mode
analysis about these points. More specifically, we shall consider
the harmonic system\begin{eqnarray*}
V(Q) & := & \V(Q_{0})+\frac{1}{2}(Q-Q_{0})^{T}w(Q-Q_{0}),\end{eqnarray*}
with the Hessian matrix $w\equiv\partial^{2}\V(Q_{0})\in\mathbb{R}^{3N\times3N}$.
According to Ehrenfest's theorem, in this case the expectation values
$\langle Q\rangle(t)$ obey the corresponding classical equations
of motion. Their solution leads us to a \emph{quadratic} eigenvalue
problem, which will be solved by linearization so as to obtain the
eigenmodes and -vectors, giving an insight into both the dynamics
and the spectrum of the problem.

\subsection{Equations of motion\label{sec:Equations-of-motion}}

Before stating the equations of motion, we are to deduce the form
of the Hessian matrix of the generalized potential $\V.$ Using the
obvious notation \begin{eqnarray*}
(\partial^{2}\V)(Q_{0}) & = & \left(\begin{array}{cc}
\frac{\partial^{2}\V}{\partial\mathbf{R}\partial\mathbf{R}} & \frac{\partial^{2}\V}{\partial\mathbf{R}\partial\mathbf{s}_{j}}\\
\frac{\partial^{2}\V}{\partial\mathbf{s}_{i}\partial\mathbf{R}} & \frac{\partial^{2}\V}{\partial\mathbf{s}_{i}\partial\mathbf{s}_{j}}\end{array}\right)_{Q_{0}}=\left(\begin{array}{cc}
w_{RR} & w_{Rj}\\
w_{iR} & w_{ij}\end{array}\right)_{i,j=1}^{N-1},\end{eqnarray*}
a lengthy calculation reveals that the symmetric matrix $w$ has the
the $3\times3$ block structure\begin{eqnarray}
w_{RR} & = & \frac{(NeB)^{2}}{M}I^{\perp}+(Ne)^{2}\left(\frac{I}{r^{3}}-\frac{3}{r^{5}}\left\{ \mathbf{R}\otimes\mathbf{R}+\frac{1}{N}\sum_{a=1}^{N}\mathbf{s}_{a}\otimes\mathbf{s}_{a}\right\} \right)\label{eq:hesse}\\
w_{Ri} & = & -Ne^{2}\frac{3}{r^{5}}\left(2\Sym\left\{ (\mathbf{s}_{i}-\mathbf{s}_{N})\otimes\mathbf{R}\right\} +\mathbf{s}_{i}\otimes\mathbf{s}_{i}-\mathbf{s}_{N}\otimes\mathbf{s}_{N}\right)\nonumber \\
w_{ij} & = & -Ne^{2}\frac{3}{r^{5}}\left(\mathbf{R\otimes R}(1+\delta_{ij})+2\Sym\left\{ (\mathbf{s}_{i}\delta_{ij}+\mathbf{s}_{N})\otimes\mathbf{R}\right\} +\mathbf{s}_{i}\otimes\mathbf{s}_{i}\delta_{ij}+\mathbf{s}_{N}\otimes\mathbf{s}_{N}\right)+\nonumber \\
 &  & Ie^{2}\left(N\frac{1+\delta_{ij}}{r^{3}}-\left\{ \delta_{ij}\sum_{{{a=1\atop (a\neq j)}}}^{N}s_{aj}^{-3}-(1-\delta_{ij})s_{ij}^{-3}+s_{iN}^{-3}+s_{jN}^{-3}++\sum_{k<N}s_{kN}^{-3}\right\} \right)+\nonumber \\
 &  & 3e^{2}\Big(\delta_{ij}\sum_{{{a=1\atop (a\neq j)}}}^{N}\frac{(\mathbf{s}_{j}-\mathbf{s}_{a})\otimes(\mathbf{s}_{j}-\mathbf{s}_{a})}{s_{ja}^{5}}+\sum_{k<N}\frac{(\mathbf{s}_{N}-\mathbf{s}_{k})\otimes(\mathbf{s}_{N}-\mathbf{s}_{k})}{s_{Nk}^{5}}+\nonumber \\
 &  & \left[\frac{(\mathbf{s}_{i}-\mathbf{s}_{N})\otimes(\mathbf{s}_{i}-\mathbf{s}_{N})}{s_{iN}^{5}}+(i\rightarrow j)\right]-(1-\delta_{ij})\frac{(\mathbf{s}_{i}-\mathbf{s}_{j})\otimes(\mathbf{s}_{i}-\mathbf{s}_{j})}{s_{ij}^{5}}\Big),\nonumber \end{eqnarray}
where $I^{\perp}\equiv\mathrm{diag}(1,1,0)$, $\mathrm{S}$ denotes
the symmetrizer: $\mathrm{S}\{ A\}=\frac{1}{2}(A+A^{T})$, and {[}$\otimes${]}
represents the tensor product. (All coordinates here of course refer
to the extremal positions.)

Despite this rather involved form, the equations of motion themselves
look fairly simple:\begin{eqnarray}
\mu_{N}\ddot{\mathbf{R}} & = & q\dot{\mathbf{R}}\times\mathbf{B}-w_{R}\cdot(Q-Q_{0})\nonumber \\
m\ddot{\mathbf{s}}_{i} & = & e\dot{\mathbf{s}}_{i}\times\mathbf{B}-\left(w_{i}-\frac{1}{N}\sum_{k}w_{k}\right)\cdot(Q-Q_{0}).\label{eq:Equations-of-motion}\end{eqnarray}
 Here a mixed notation was employed for clarity: $w_{a}\equiv(w_{aR},\, w_{a1},\dots,w_{a,N-1})\in\mathbb{R}^{3\times3N}$
is the $a$-th block row of $w$. It is inviting to interpret these
equations as the motion of effective particles with masses $\mu_{N},\, m$
and charges $q,\, e$, respectively, in a magnetic field, coupled
via linear forces. 

To solve the equations of motion, let us first write the above system
in a more compact way in terms of the \emph{displacements} $v(t):=Q(t)-Q_{0}$,\begin{equation}
\ddot{v}=\omega\cdot\dot{v}+A\cdot v.\label{eq:eom-for-v}\end{equation}
Here the antisymmetric \emph{cyclotron matrix $\omega$} and the \emph{harmonic
matrix} $A$, respectively, have been introduced:\begin{eqnarray*}
\omega & = & \frac{qB}{\mu}\varepsilon\oplus\frac{eB}{m}\varepsilon\oplus\dots\oplus\frac{eB}{m}\varepsilon\in\mathbb{R}^{3N\times3N}\\
A & = & -\left(\begin{array}{c}
w_{R}/\mu\\
\left(w_{i}-\frac{1}{N}\sum_{k}w_{k}\right)/m\end{array}\right)_{i=1,\dots,N-1}\negthickspace\in\mathbb{R}^{3N\times3N},\end{eqnarray*}
where $(\varepsilon)_{ij}:=\varepsilon_{ij3}$ is connected with the
Levi-Civita tensor and {[}$\oplus${]} stands for the direct sum.
The solution of this $3N$-dimensional second-order system of differential
equations (ODE) is given by the span\begin{equation}
v(t)=\sum_{\rho=1}^{6N}(v_{\rho}e^{\gamma_{\rho}t})c_{\rho}\qquad(v_{\rho}\in\mathbb{C}^{3N};\,\gamma_{\rho},c_{\rho}\in\mathbb{C}),\label{eq:NM_sum}\end{equation}
 fulfilling the \emph{quadratic} eigenvalue equation\begin{equation}
(\gamma_{\rho}^{2}I-\gamma_{\rho}\omega-A)v_{\rho}=0.\label{eq:QEP}\end{equation}
In this respect, our stability analysis amounts to finding the complex
\emph{eigenvalues} $\gamma_{\rho}=:\Gamma_{\rho}+i\Omega_{\rho}$
(whose imaginary parts are frequencies of a vibration about a stable
point, and whose real part corresponds to an instability), and the
\emph{eigenvectors} $v_{\rho}$.

The above quadratic eigenvalue problem is solved via \emph{linearization},
which is entirely equivalent to reducing a second-order ODE to a first-order
system. In this fashion one obtains a standard \emph{linear} eigenvalue
problem \begin{equation}
\ca u=\gamma u,\label{eq:QEP-linear}\end{equation}
with the \emph{linearization matrix}\[
\ca=\left(\begin{array}{cc}
0 & I\\
A & \omega\end{array}\right)\in\mathbb{R}^{6N\times6N}.\]
 Before we present the results, let us state two general properties
of the solutions that can be obtained ahead of a computation:

\begin{enumerate}
\item For any eigenpair $(\gamma,v)$ of (\ref{eq:QEP}) such that \textrm{Im$\gamma\neq0$} 
, there is another pair $(\gamma^{*},v^{*})$. 
\item Let $\omv:=v^{\dagger}\omega v$ denote the Rayleigh quotient for
an arbitrary matrix $\omega$. Assuming $v$ to be a normalized eigenvector,
then for any mode the eigenvalue is given by\[
\gamma^{\pm}=\frac{1}{2}\left(\omv\pm\sqrt{\omv^{2}+4\av}\right),\]
where in general only one of the roots is in the spectrum.
\end{enumerate}
Firstly, (1.) implies that only \emph{one} out of two modes is relevant;
hence it is widely legitimate to treat the $6N$ modes as effectively
$3N$. (2.) tells us where in the complex plane we can expect the
eigenmodes to lie. If the problem is \emph{elliptic}, that is\begin{equation}
\av<-\left(\frac{\omv}{2}\right)^{2}\;\forall v,\label{eq:crit-elliptic}\end{equation}
 then the solution is imaginary $\gamma=i\Omega$ (since $\omega$
is antisymmetric).

\subsection{Results and discussion\label{sub:Results-and-discussion}}

\subsubsection{General Classification\label{sub:General-Classification}}

\begin{figure}
\begin{center}\includegraphics[%
  width=11cm,
  keepaspectratio]{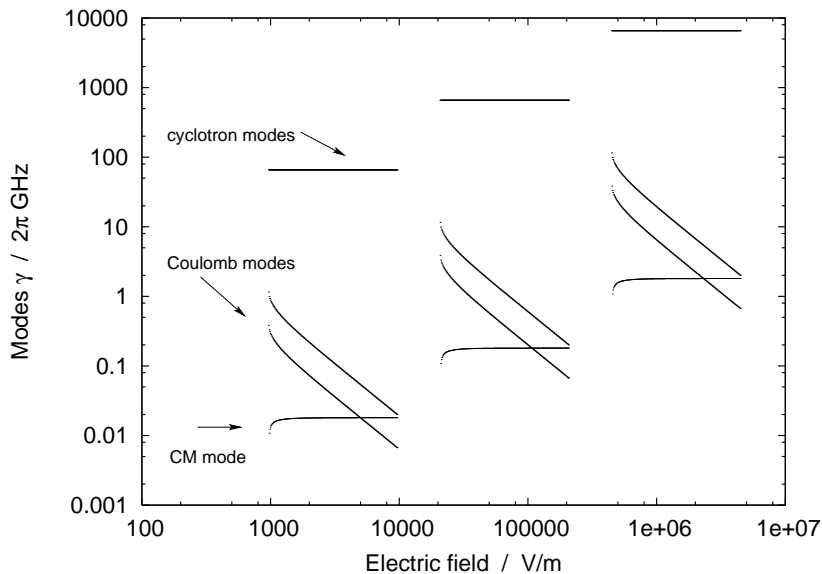}\end{center}

\caption{Eigenmodes $\{\gamma_{\rho}\}_{\rho}$ for $N=2$ electrons as a
function of the electric field $E\equiv BK/M$ over the range $K/K_{cr}\in[1,10]$
(all modes are imaginary). The three plots refer to the magnetic fields
$B=10^{-5},10^{-4},10^{-3}$ (from left to right). Each top horizontal
line represents two almost degenerate cyclotron modes. Below, the
two Coulomb modes fall off quickly and intersect the CM mode (the
nearly horizontal line about four orders below the cyclotron modes).\label{cap:NMA}}
\end{figure}
\begin{figure}
\begin{center}\subfigure[Case of three electrons]{\includegraphics[%
  width=8cm,
  keepaspectratio]{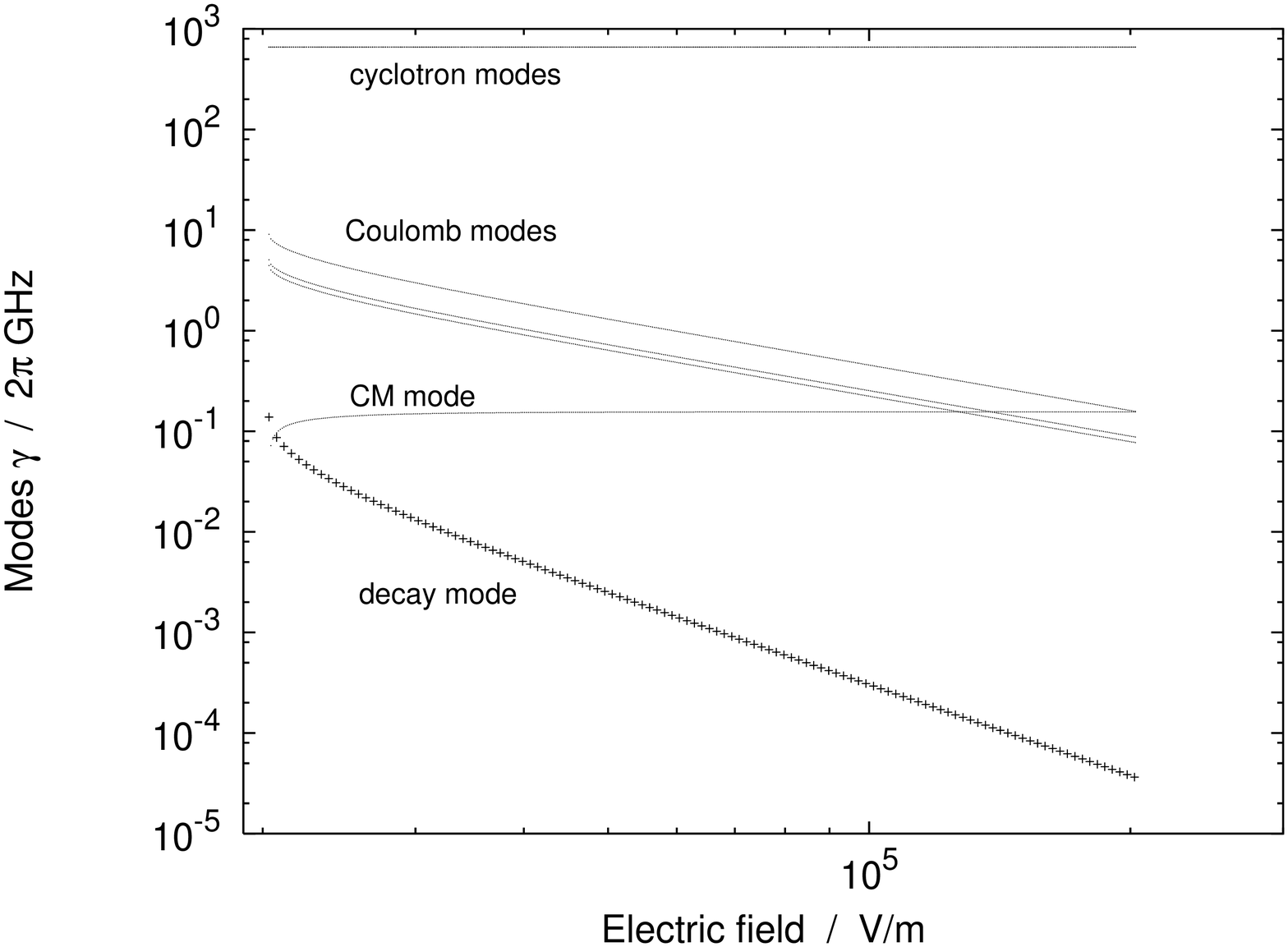}}\subfigure[Case of four electrons]{\includegraphics[%
  width=8cm,
  keepaspectratio]{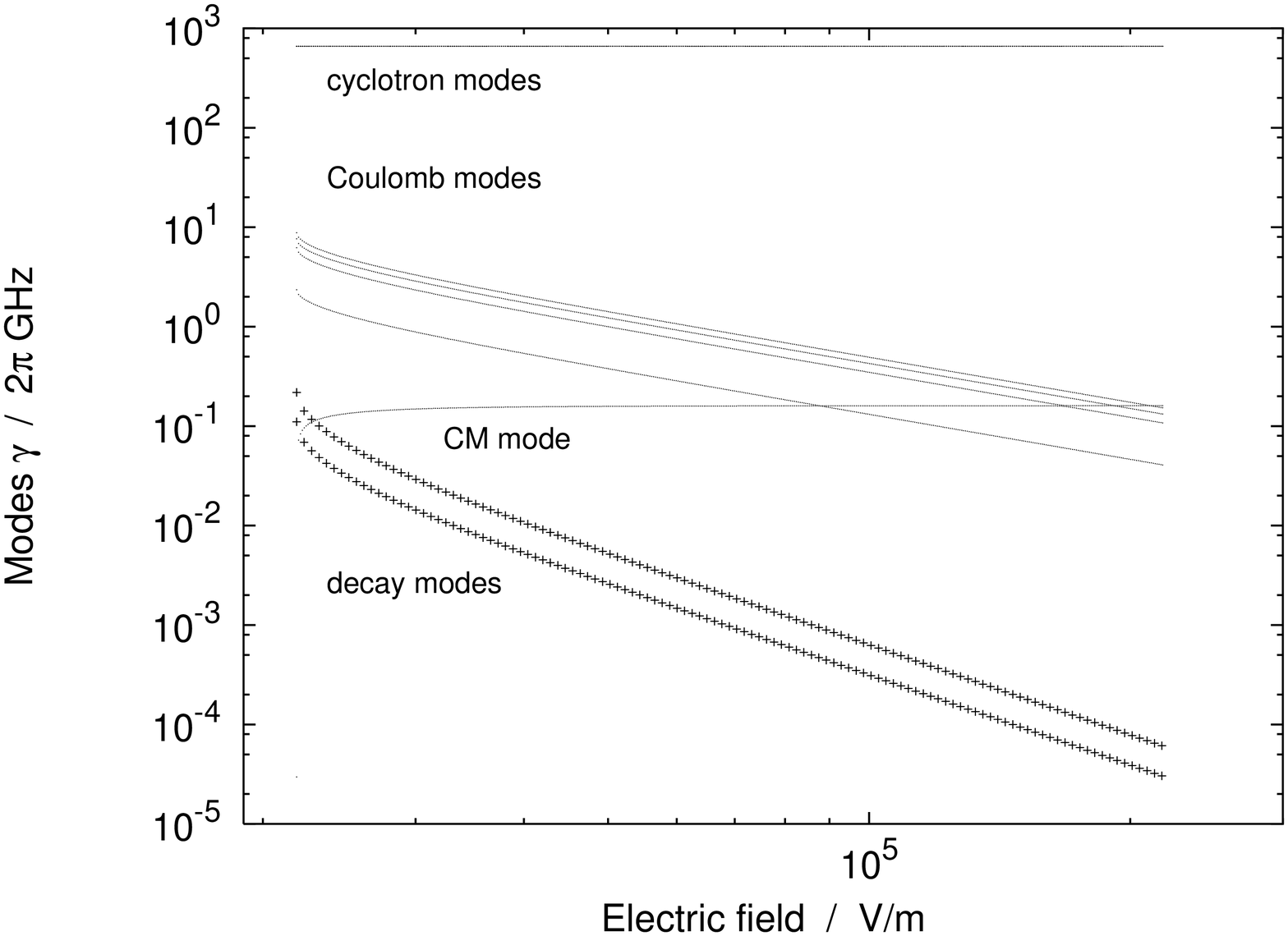}}\vspace{-0.5cm}\end{center}

\begin{center}\subfigure[Case of five electrons]{\includegraphics[%
  width=8cm,
  keepaspectratio]{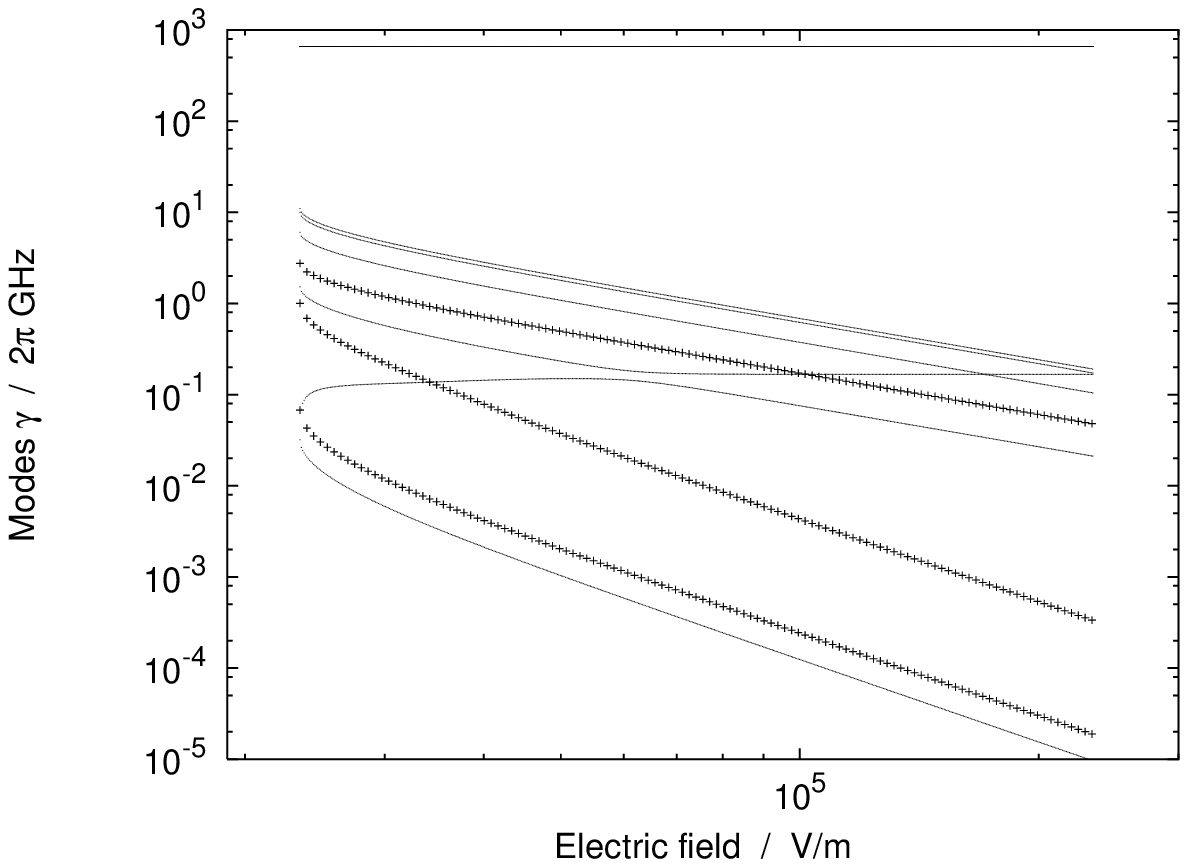}}\subfigure[Case of ten electrons]{\includegraphics[%
  width=8cm,
  keepaspectratio]{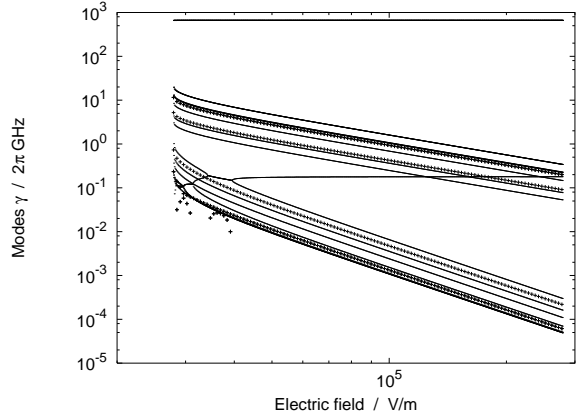}}\vspace{-0.5cm}\end{center}

\caption{Eigenmodes $\gamma$ for the cases $N=3$ (a), $N=4$ (b), $N=5$
(c) and $N=10$ (d), as a function of the electric field $E\equiv BK/M$,
plotted over the range $K/K_{cr}\in[1,10]$. The magnetic field is
$B=10^{-4}$. Imaginary parts appear as solid lines, while points
(+) are used for the real parts. The individual modes are assigned
just as for Fig. \ref{cap:NMA}, with an extra class of very small
\emph{decay modes}, which fall off twice as fast as the Coulomb lines.
\label{cap:NMA3-10}}
\end{figure}
Let us first have a look at the behavior of the modes $\{\gamma_{\rho}(K)\}_{\rho}$
for different $N$ at fixed $B$ (Figures \ref{cap:NMA}, \ref{cap:NMA3-10}).
 Even though the patterns become increasingly rich and involved with
higher $N$, one can see a distinction between the modes regarding
their behavior as a function of $K$ (and $B$), their order of magnitude
and, more generally, their location in the complex plane.

An analysis of the associated eigenvectors and the criterion for ellipticity,
(\ref{eq:crit-elliptic}), will also allow us to illuminate the modes'
origins. This suggests the following classification \textbf{}of the
$3N$ \emph{eigenmodes}:

\begin{itemize}
\item $N$ so-called \emph{cyclotron modes} corresponding to the cyclotron
motion of the effective particles. Their values are exclusively on
the scale of the cyclotron frequencies, $\gamma=i\Omega\sim i\frac{|e|B}{m}$
\item $N$ \emph{Coulomb} \emph{modes} corresponding to the inter-Coulombic
motion, with a main contribution from the harmonic matrix $A$. They
fall off quickly with $K$, since the decentering $X_{0}$ increases,
and so the Coulomb interaction becomes less effective.
\item 1 \emph{CM mode}, roughly reflecting the cyclotron motion of the center
of mass , $\Omega\sim\frac{NeB}{M}$ (stemming mostly from the CM-kinetic
energy)
\item 1 \emph{zero mode} $(\gamma_{\rho}=0)$ that stems from the rotational
invariance of the saddle point with respect to the circular configuration
(see Eq. \ref{eq:ass-2})
\item $N-2$ modes that will be referred to as \emph{decay modes} in recognition
of the fact that they \emph{}are predominantly real\emph{.} They are
neither directly related to the cyclotron motion nor to the spectrum
of $A$, and their slope is twice as steep as that of the Coulomb
modes.
\end{itemize}
\begin{figure}
\begin{center}\includegraphics[%
  width=7cm]{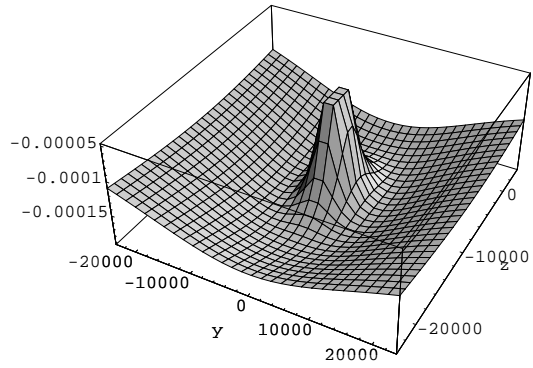}\end{center}

\caption{Visualization of the zero mode: plot of the generalized potential
$\mathcal{V}$ for $N=2$ close to the extremum, cut along the $yz$
coordinates of $\mathbf{s}$ ($K/K_{cr}=2;B=10^{-4}$). There is a
circle that continuously connects all minima, referred to as {}`stationary
circle'. \label{cap:zero-mode}}
\end{figure}
 The cyclotron modes roughly pertain to gyrations perpendicular to
$\mathbf{B}$, while the Coulomb motion is predominantly parallel
to $\mathbf{B}$. The CM mode is primarily a gyrational mode---reflecting
the nucleus motion---, but it also exhibits Coulomb-type contributions.
The zero mode---to be analyzed in more detail below---corresponds
to the rotation along the stationary circle identifiable in Fig. \ref{cap:zero-mode}.
Lastly, the {}`decay modes' reveal no striking underlying structure.

There are a few subtleties that go beyond the categorization suggested
above. A closer look at, for instance, the case of $N=10$ electrons
reveals certain interactions among the different modes. Their principal
causes are crossings between the center-of-mass mode and the decay
modes (resulting in some striking deformations of the usual line pattern),
and equally avoided crossings of the CM mode with at least some of
the Coulomb modes. This may be taken as a hint at the different symmetry
relations among the Coulombic and the decay modes.

\subsubsection{Influence of the rotational invariance $\Delta$\label{sub:rotation}}

By construction, the stationary character was not affected by a common
rotation of the relative coordinates $\mathbf{s}_{a}$ by an angle
$\Delta$ on the circle: \begin{eqnarray}
\mathbf{s}_{a} & \mapsto & D_{\Delta}\mathbf{s}_{a}\quad\forall a,\label{eq:s_Delta}\\
\phi_{a} & \mapsto & \phi_{a}+\Delta\nonumber \end{eqnarray}
 where $D_{\Delta}\in1\oplus\mathrm{SO}(2)$ denotes that rotation
about the $\xib$ axis. Nonetheless, the Hessian matrix $\partial^{2}\V\big|_{Q_{0}}\equiv w$
is affected---and thus the dynamics differs. To see this, note that
the block matrices transform like\begin{eqnarray*}
w_{RR} & \stackrel{D_{\Delta}}{\longmapsto} & w_{RR}\;(N>2)\\
w_{aj} & \stackrel{D_{\Delta}}{\longmapsto} & D_{\Delta}w_{aj}D_{\Delta}^{-1}.\end{eqnarray*}
The spectra of both the Coulombic Hessian $(\partial^{2}V)_{0}$ and
the magnetic contributions (including $\omega$) are separately invariant
under these rotations. Still, their conflicting symmetries guarantee
that the dynamics will vary with $\Delta$.

\begin{figure}
\begin{center}\subfigure[Case of two electrons]{\includegraphics[%
  width=11cm]{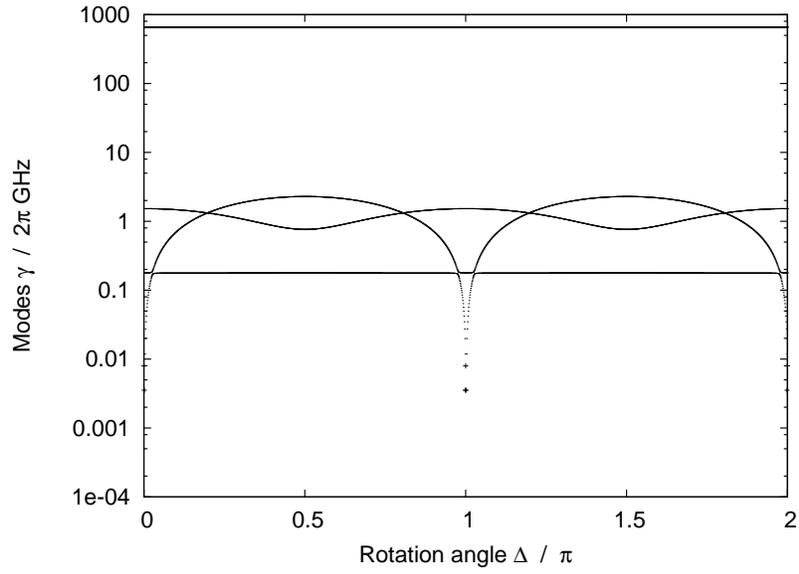}}\vspace{-0.5cm}\end{center}

\begin{center}\subfigure[Case of three electrons]{\includegraphics[%
  width=11cm]{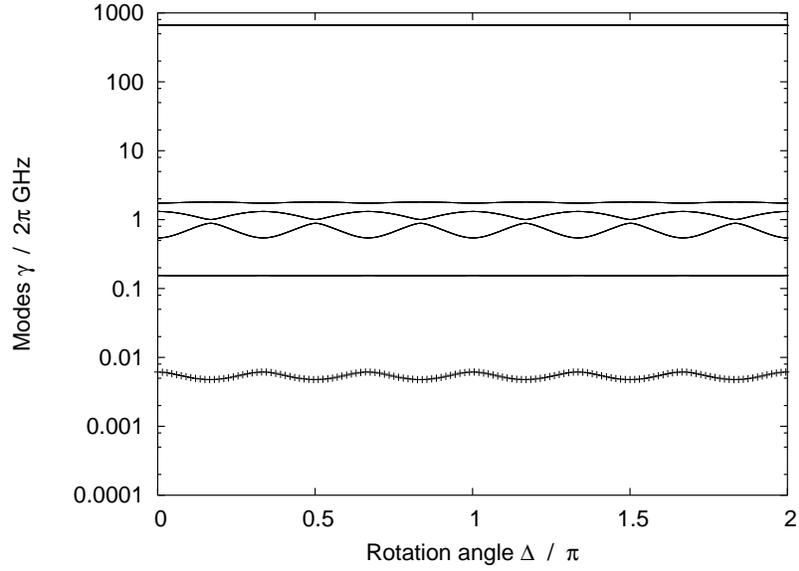}}\vspace{-0.5cm}\end{center}

\caption{Upon rotating the circular configuration by an angle $\Delta$, the
spectrum is changed in part ($B=10^{-4},\, K=2K_{cr}$). The cyclotronic
modes $(658\,\mathrm{GHz})$ and CM mode $(0.17\,\mathrm{GHz})$ remain
constant. The Coulomb modes and, for $N=3$, the decay mode ($\sim10^{-2}\mathrm{GHz}$)
show some periodic modulation. (for symbol coding, see Fig. \ref{cap:NMA3-10})
\label{cap:N2/3-delta}}
\end{figure}

In order to see this dependency, a semi-log plot of the modes $\{\gamma_{\rho}(\Delta)\}$
on the maximal interval $[0,2\pi)$ is recorded for different $N$
(Fig. \ref{cap:N2/3-delta}). As before, $B=10^{-4}$ and $K=2K_{cr}$
are fixed. The plots display an obvious symmetry, as rotating by $\Phi_{N}=2\pi/N$
gives an indistinguishable setup. The cyclotronic modes as well as
the affiliated CM mode are essentially $\Delta$-independent. By contrast,
the Coulomb modes and the decay modes show a pronounced periodic change.
In addition, one can see avoided crossings in some cases between the
CM mode and the lowest Coulomb mode.

\begin{figure}
\begin{center}\includegraphics[%
  width=11cm]{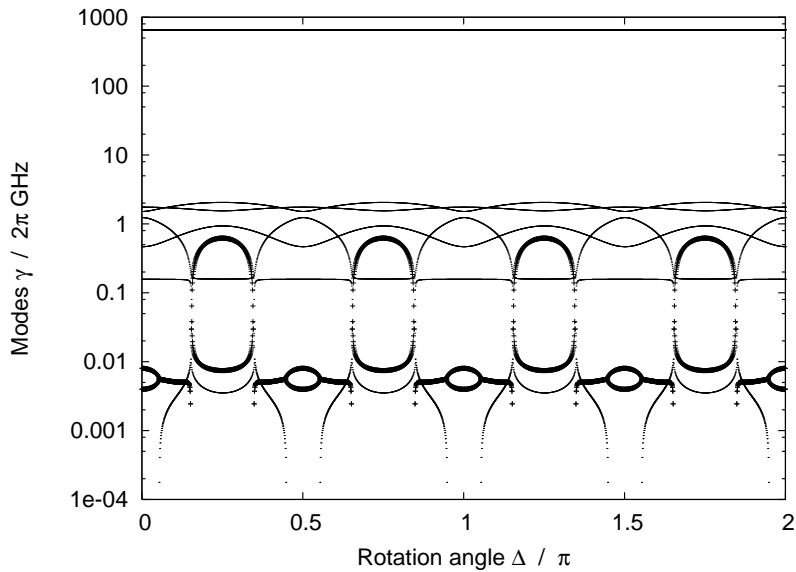}\vspace{-0.5cm}\end{center}

\caption{Spectrum for $N=4$ depending on the rotation angle $\Delta$ ($B=10^{-4},\, K=2K_{cr}$);
see text. \label{cap:N4-delta}}
\end{figure}
For the case $N=2$ (Figure \ref{cap:N2/3-delta}a), which was found
to be locally stable in the vertical configuration (\cite{Schm01},
see also \ref{sub:Results-and-discussion}), there is some exceptional
behavior whenever $\Delta$ is close to $k\cdot\pi$. The lower Coulomb
mode has an avoided crossing with the almost constant CM mode before
it experiences a singularity on the logarithmic scale, along the way
turning real. In this respect, the local stability we found for $N=2$
is applicable only outside the singular \emph{horizontal} configuration
$\Delta=0$.

While the case of three electrons ($N=3$: Fig. \ref{cap:N2/3-delta}b)
reveals no striking effects, the spectrum becomes much richer when
we add another electron ($N=4$: Fig. \ref{cap:N4-delta}). The four
Coulomb modes as well as two more decay modes reveal an intriguing
behavior. There is an avoided crossing between the CM mode and the
lowest Coulomb mode, which turns into an unstable \emph{real} mode
in an intermediary region. Also, one can get a sense of the interplay
between the two decay modes.

\section{Wave-packet dynamical study of the two-electron case\label{sec:Wave-packet-dynamical-study}}

The extensive analysis of the normal modes of $N$-electron giant
dipole states in the preceding section provided evidence that the
two-electron case is locally stable, apart from the singular horizontal
configuration. We now turn to a numerical study of that system. 

It cannot be over-emphasized that a six-dimensional resonance study
is at the frontier of what is currently possible and requires a careful
choice of the computational approach. This applies especially in view
of the fact that the system is governed by dramatically different
time scales. Therefore we adopted the Multi-Configuration Time-Dependent
Hartree \noun{(mctdh)} method \cite{mey90:73,bec00:1,mey03:251,mctdh:package},
which is known for its outstanding efficiency in high-dimensional
applications. To be self-contained, a concise introduction to this
tool is presented in Subsection \ref{sec:method}. Its basic idea
is to solve the time-dependent Schrödinger equation in a small but
time-dependent basis related to Hartree products. It has to be stressed
that this method is designed for distinguishable particles. Applying
it to a fermionic system like ours finds its sole justification in
the fact that the spatial separation between the electrons is so large
that they are virtually distinguishable.

After some remarks on the scales of the system in Subsec. \ref{sec:Preliminary-considerations},
the method will be applied in Subsec. \ref{sec:MCTDH-Results-and-discussion}.
The focus is on stability aspects, i.e., the propagation of wave packets,
but we also investigate the spectra as well as the influence of the
rotational invariance discussed in the previous section.

\subsection{Computational method: \noun{mctdh} \label{sec:method}}

The principal goal of \noun{mctdh} is to solve the time-dependent
Schr\"{o}dinger equation as an initial-value problem: \begin{equation}
\left\{ \begin{array}{c}
i\dot{\Psi}=H\Psi\\
\Psi(Q,0)=\Psi_{0}(Q)\end{array}\right..\label{eq:TDSE}\end{equation}
The \emph{standard method} (see \cite{mey03:251} and references therein)
now approaches the problem by expanding the solution in terms of time-independent
\emph{primitive basis functions}. Unfortunately, it exhibits a drastic
exponential scaling with respect to the number of degrees of freedom
$f$. Rather than using a \emph{large} time-independent basis, the
\emph{Multi-Configuration Time-Dependent Hartree} method (\noun{mctdh})
employs a time-dependent set, which is more flexible and thus \emph{smaller}.
The ansatz for the wave function now reads \[
\Psi(Q,t)=\sum_{j_{1}=1}^{n_{1}}\ldots\sum_{j_{f}=1}^{n_{f}}A_{j_{1}\ldots j_{f}}(t)\prod_{\kappa=1}^{f}\varphi_{j_{\kappa}}^{(\kappa)}(Q_{\kappa},t)\equiv\sum_{|J|\le|n|}A_{J}(t)\Phi_{J}(Q,t),\]
using a convenient multi-index notation for the configurations, $J=(j_{1}\dots j_{f})$,
where $f$ denotes the number of degrees of freedom. The (unknown)
\emph{single-particle functions} $\varphi_{j_{\kappa}}^{(\kappa)}$
are in turn represented in a primitive basis. Note that both the coefficients
$A_{J}$ and the Hartree products $\Phi_{J}$ carry a time-dependence;
hence the uniqueness of solutions can only be ensured by adding certain
constraints on the single-particle functions \cite{bec97:113,bec00:1}.
Using the Dirac-Frenkel variational principle \cite{dir30:376}, one
can derive equations of motion for both $A_{J,}\,\Phi_{J}$ \cite{bec97:113,bec00:1}.
Integrating these ODE systems allows us to obtain the time evolution
of the system, $\Psi(\cdot,t)=\sum_{J}A_{J}(t)\Phi_{J}(\cdot,t)$,
which lies at the heart of wave-packet propagation.

The Heidelberg \noun{mctdh} package \cite{mctdh:package}, which
we used, incorporates a few extensions to this basic concept:

\begin{itemize}
\item \emph{Product representation of the potential}: in order to circumvent
multi-dimensional integration in computing the matrix elements, \noun{mctdh}
makes the requirement\begin{equation}
H=\sum_{r=1}^{s}c_{r}\prod_{\kappa=1}^{f}h_{r}^{(\kappa)}\,,\label{hresidual}\end{equation}
enforcing that the \emph{numerical} Hamiltonian be written as a sum
of products of one-particle operators (separable form). The non-separable
part of the Hamiltonian thus has to be fitted to product form ahead
of a computation.
\item \emph{Mode combination}: in practice, one combines several (say $d=2,3$)
degrees of freedom to a $d$-particle rather than a \emph{one}-particle
function. This approach alleviates the bad numerical scaling for high-dimensional
systems.
\item \emph{Relaxation} \cite{kos86:223}: \noun{mctdh} provides a way
to not only propagate a wave packet, but also to obtain the lowest
eigenstates of the (discretized) system. The underlying idea is to
propagate some wave function by the non-unitary time-evolution operator
$e^{-H\tau}$ (\emph{propagation in imaginary time}.) In practice,
one relies on a more sophisticated scheme termed \emph{improved relaxation}.
Here $\langle\Psi|H-E|\Psi\rangle$ is minimized with respect to both
the coefficients $A_{J}$ and the configurations $\Phi_{J}$. The
equations of motion thus obtained are then solved iteratively by first
solving for $A_{J}(t)$ (by diagonalization of $(\langle\Phi_{J}|H\Phi_{K}\rangle)$
with fixed $\Phi_{J}$) and then propagating $\Phi_{J}$ in imaginary
time over a short period. The cycle will then be repeated.
\item \emph{Spectrum}: one can also extract information on the spectrum
by computing the auto-correlation function $c(t):=\langle\Psi_{0}\mid e^{-iHt}\Psi_{0}\rangle.$
In the case of a purely discrete spectrum, its Fourier transform reads\begin{equation}
\tilde{c}(E)=2\pi\langle\Psi_{0}\mid\delta(H-E)\Psi_{0}\rangle=2\pi\sum_{J}\delta(E_{J}-E)|c_{J}|^{2},\label{eq:auto-corr-F}\end{equation}
that is, all eigen-energies give a peak-type contribution, weighted
according to the overlap $c_{J}$ of $\Psi_{0}$ with the $J$-th
eigenstate.
\end{itemize}

\subsection{Application of \noun{mctdh}\label{sec:Preliminary-considerations}}

Let us point out how the computational method can be applied to our
problem. There are essentially two highly conflicting types of motion
and corresponding scales: \emph{perpendicular} to the magnetic field,
we have the magnetic length\begin{eqnarray*}
R_{B} & = & \frac{1}{\sqrt{\mu\omega_{B}}}\simeq\frac{1}{\sqrt{2B}},\end{eqnarray*}
employing atomic units. To be specific, for a laboratory field strength
$B=10^{-4}$, the length scale will be on the order of $10^{2}$ a.u.
As for the \emph{parallel (Coulomb) motion}---expected to take place
roughly in a harmonic potential---we can use the usual harmonic-oscillator
lengths $Z_{0}=1/\sqrt{\mu\omega_{Z}}$, for some {}`typical' frequency
$\omega_{Z}$. These are roughly on the order of $10^{3}$ a.u. (Note
that the higher $K$---or $E$--- becomes, the smaller will be the
Coulomb frequencies, hence the wave packet will be spread far more
over the grid.)

Analogously, the anticipated time scales are\[
T_{B}=\frac{2\pi}{\omega_{B}}\stackrel{B=10^{-4}}{\simeq}1.5\,\mathrm{ps}\]
for the cyclotronic motion. By contrast, the characteristic times
parallel to \textbf{B} are strongly $K$-dependent, and they vary
between about $1,500\mathrm{ps}-15,000\mathrm{ps}$. For comparison,
we mention that the associate time scale for the CM mode is asymptotically
$T_{CM}\simeq5,000\,\mathrm{ps}$.

For strong fields, the relevant motion will thus certainly be determined
by the magnetic field---at least near the stationary points we are
interested in. This would suggest to introduce cylindrical coordinates
about the extremum, carrying out our study in the basis set for the
\emph{Landau orbitals} \cite{landau} $R_{nm}(\rho)e^{im\varphi}$
perpendicular to $\mathbf{B}$, and harmonic-oscillator functions
$\phi_{n_{z}}(z)$ parallel to the field. Since the former set was
not available in the \noun{mctdh} package, we resorted to Cartesian
coordinates for the displacements from the extremum $Q_{0}$ as a
pragmatic solution:\[
\mathbf{s}\mapsto\mathbf{s}-\mathbf{s}_{0}\equiv(x,y,z),\quad\mathbf{R}\mapsto\mathbf{R}-\mathbf{R}_{0}\equiv(X,Y,Z).\]
 The Hamiltonian (\ref{eq:kinetic_ECM}-\ref{eq:potential_ECM}) for
our case, centered about the extremum $Q_{0}$ by the generalized
translation operators $\exp i(\mathbf{k}_{R}\cdot\mathbf{R}_{0}+\mathbf{k}_{s}\cdot\mathbf{s}_{0})$,
reads:\begin{eqnarray*}
\mathcal{T} & = & -\frac{1}{2\mu}\nabla_{R}^{2}+\frac{1}{8}\mu\omega_{R}^{2}(X^{2}+Y^{2})+\frac{\omega_{R}}{2}L_{Z}+\\
 &  & -\frac{1}{4m}\nabla_{s}^{2}+\frac{1}{4}m\omega_{s}^{2}(x^{2}+y^{2})+\frac{\omega_{s}}{2}\ell_{z}\\
\mathcal{V} & = & \frac{1}{2M}\left((2eB)^{2}\left[X^{2}+Y^{2}\right]+2X\cdot(X_{0}(2eB)^{2}-2eBK')\right)+V_{0}+V_{1}+V_{2}+\mathcal{C}\end{eqnarray*}
where $\omega_{R}=|q|B/\mu,\,\omega_{s}=|e|B/m$. $\mathcal{C}=\frac{1}{2M}(\mathbf{K'}-2e\mathbf{B}\times\mathbf{R}_{0})^{2}$
only shifts the energy by a constant and will be ignored hereafter,
and $r_{0}=\tilde{\alpha}|X_{0}|,\, s_{0}=r_{0}/\alpha,\,\phi_{0}=\pi+\Delta$
denote the extremal configuration. Moreover, $V_{0}$ stands for the
electron-electron repulsion, whereas $V_{1/2}$ denotes the attractive
potentials between the nucleus and electron \#1 and \#2, respectively.
Both $\mathcal{T}$ and the {}`CM-kinetic energy' are already in
product form (\ref{hresidual}), whereas the Coulomb terms have to
be fitted.

\subsection{Results and discussion\label{sec:MCTDH-Results-and-discussion}}

In our calculations, we were chiefly interested in two questions:
the stability of the giant dipole states, and their spectral properties.
The first aspect was studied using propagation of wave packets initially
localized at the extremum. Emphasis was placed on initial states close
to the assumed {}`ground states'. Moreover, following the evolution
of wave packets with an initial displacement from the extremum, we
tested the robustness of the resonances. For comparison, the eigenvectors
of the Hamiltonian will be considered using the relaxation method.

The parameter set was restricted to the following values. $B=10^{-4}$
was fixed, because changing the magnetic field strength essentially
only affects the overall scales via the critical pseudomomentum and
is not expected to make for completely new behavior. The pseudomomenta
are $K/K_{cr}\in\{1.1,\,2.0,\,10.0\}$ (we will drop the prime from
now on), which account for the cases of a very shallow \emph{outer
well}, the medium range and the very deep outer well (the high-$K$
regime).

For simplicity, the influence of the rotational invariance will be
scrutinized \emph{a posteriori}, using the characteristic values $\phi_{0}=\frac{\pi}{2},\,\frac{3\pi}{4},\,\pi$.
These refer to the vertical configuration ($\Delta=-\frac{\pi}{2}$)
first studied in \cite{Schm01}, a {}`diagonal' one $\Delta=-\frac{\pi}{4}$,
and the horizontal configuration $\Delta=0$, which we expect to have
an instability according to the analysis in \ref{sub:rotation}.

The propagation times were chosen in the regime of $5-10\times10^{4}\,\mathrm{ps}$.
This includes many periods of the Coulomb modes and several $10,000$
periods of the rapid cyclotron motion.

\subsubsection{Stability analysis}

The approach used in this subsection is to study the propagation of
a 6D harmonic-oscillator wave packet centered within the well. The
oscillator parameters are adapted so as to match the borderline ground
states, i.e., the cyclotron frequencies $\omega_{R},\,\omega_{s}$
for $XY(xy)$, and a typical Coulomb frequency $\omega_{Z},\,\omega_{z}$
parallel to the magnetic field. To test the robustness of the potential
resonance, in some cases \emph{displacements} of $\exv{Z}_{0}=\exv{z}_{0}=2000$~a.u.
are also applied to the initial wave function. (This value is arbitrary
but moderate compared to the total decentering.) Also, \emph{relaxations}
(starting from the original, undisplaced wave packets above) are carried
out to see how the lowest-energy configuration found with this algorithm
differs from our initial guess. The analysis mostly focuses on the
\emph{reduced} densities $\rho_{\kappa}$ (1D or 2D) obtained by integrating
out all but the $\kappa$-th degree(s) of freedom. Where appropriate,
the wave packets' centers and widths $\langle q_{\kappa}\rangle(t),\,\Delta q_{\kappa}(t)$
are shown.

\paragraph{The case $K=1.1K_{cr}$:\label{par:K1.1}}

Very close to the critical point, $K_{cr}$, the observed motion displays
an instability in some degrees of freedom. 

\begin{figure}
\begin{center}\subfigure[]{\includegraphics[%
  width=7.5cm,
  keepaspectratio]{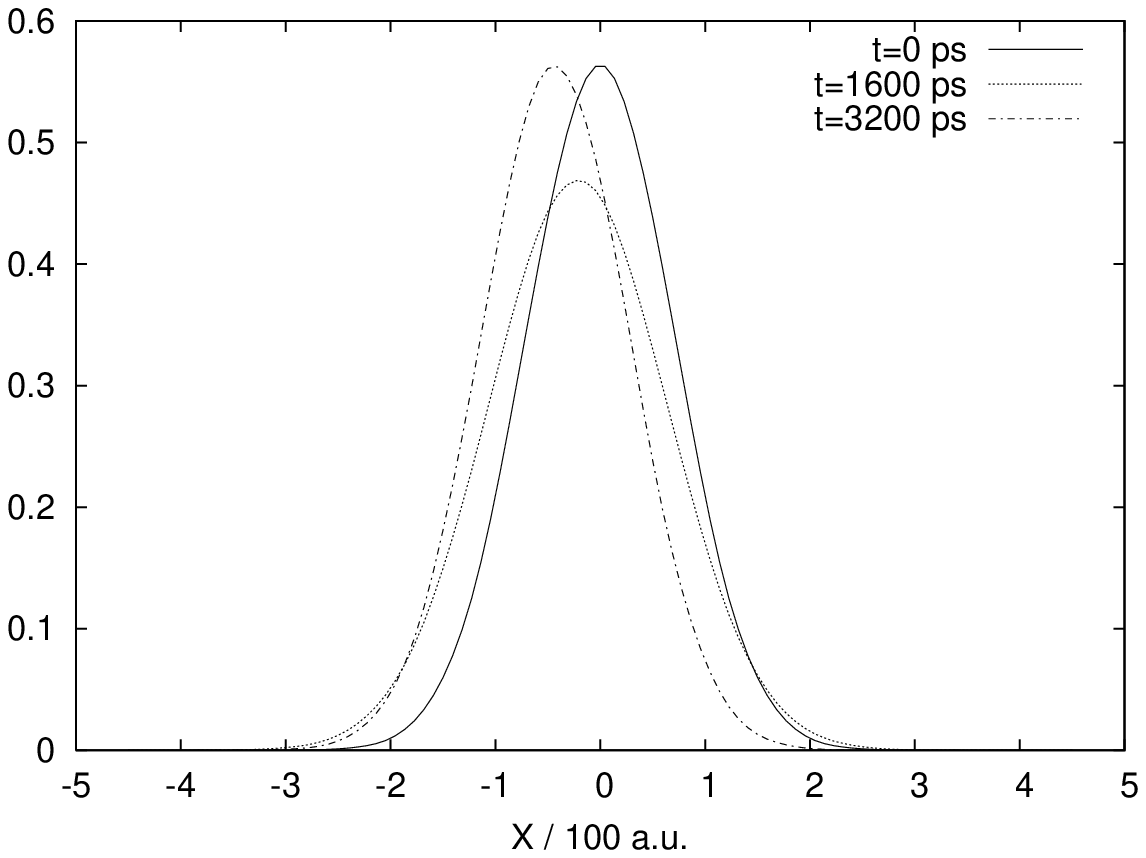}}\subfigure[]{\includegraphics[%
  width=8cm,
  keepaspectratio]{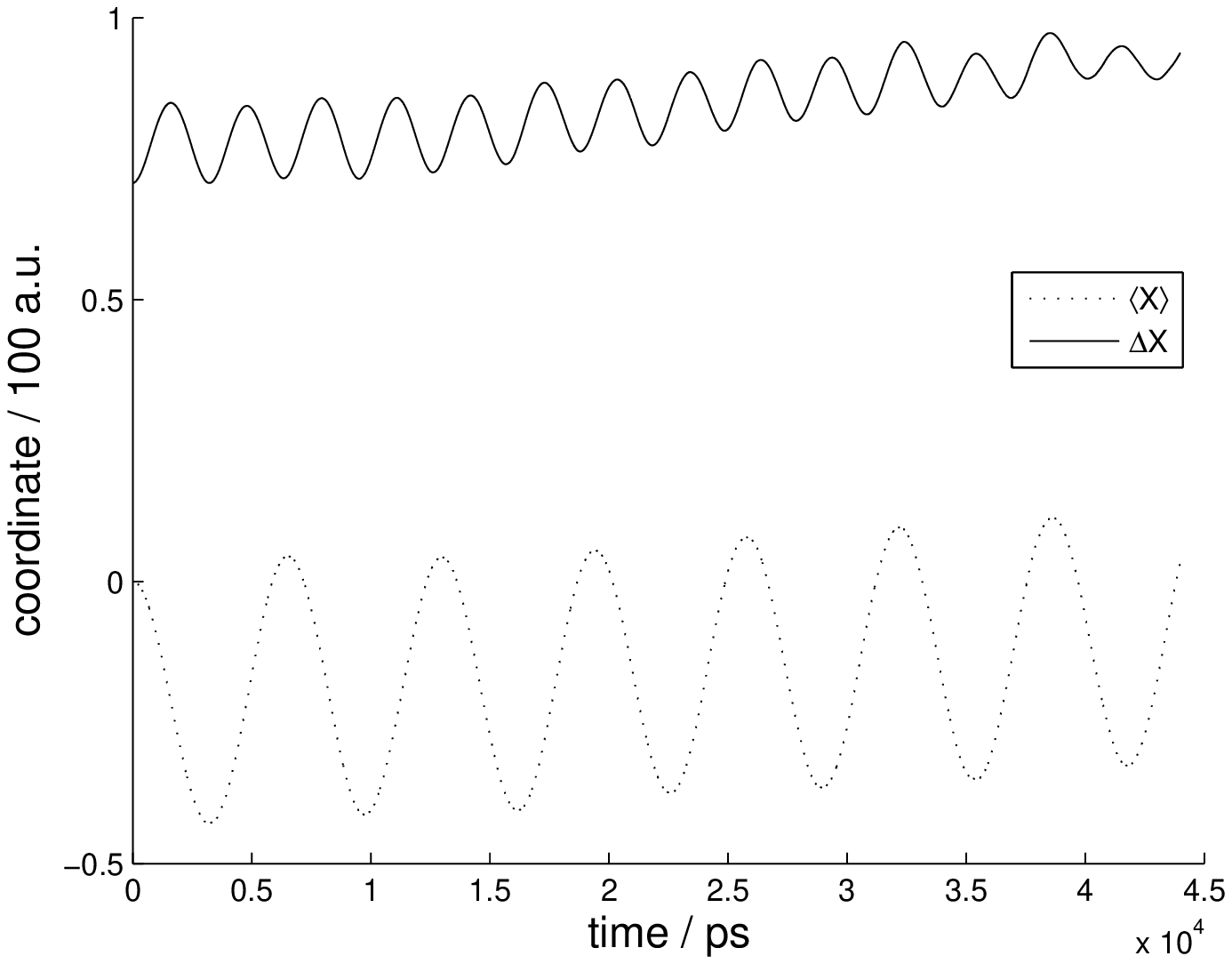}}\vspace{-0.5cm}\end{center}

\caption{The motion in $X$ in the case $K/K_{cr}=1.1$ \protect \\
(a) Snapshots of the motion $\rho_{X}$, reflecting oscillations of
both center and shape of the wave packet \protect \\
(b) The two oscillations as documented in $\exv{X}(t),\,\Delta X(t)$
\label{cap:K1.1X_snap}}
\end{figure}
Both $X$ and $Y$ revealed weak oscillations of two types. As these
are generic, they shall be inspected in more detail on the example
of the $X$ degree of freedom. Three representative snapshots (Figure
\ref{cap:K1.1X_snap}a) were taken to illustrate this effect. The
nature of the oscillations can be explored more easily in Figure \ref{cap:K1.1X_snap}(b).
To begin with, the center carries out a tiny oscillation about $X\simeq-20$
a.u. with a period of about $6,400\,\mathrm{ps}$ (a snapshot was
taken at its first turning point at $3,200\,\mathrm{ps}$). This very
slow oscillation superimposes what we will call the \emph{shape oscillations},
materializing in the fluctuation of the width $\Delta X(t)$. That
oscillation is about twice as fast, and it reaches its maximum width
for the first time at $1,600\,\mathrm{ps}$ (Fig. \ref{cap:K1.1X_snap}a).
The relative stability in these two directions is readily explained
in terms of the generalized potential. Apart from a stabilization
by the $\mathbf{B}$ field, they experience an additional confinement
via the {}`CM-kinetic energy' (\ref{eq:CM-Ekin}). However, it is
evident that the packet broadens over time, thus slowly delocalizing
(Fig. \ref{cap:K1.1X_snap}b). 

\begin{figure}
\begin{center}\subfigure[]{\includegraphics[%
  width=7cm,
  keepaspectratio]{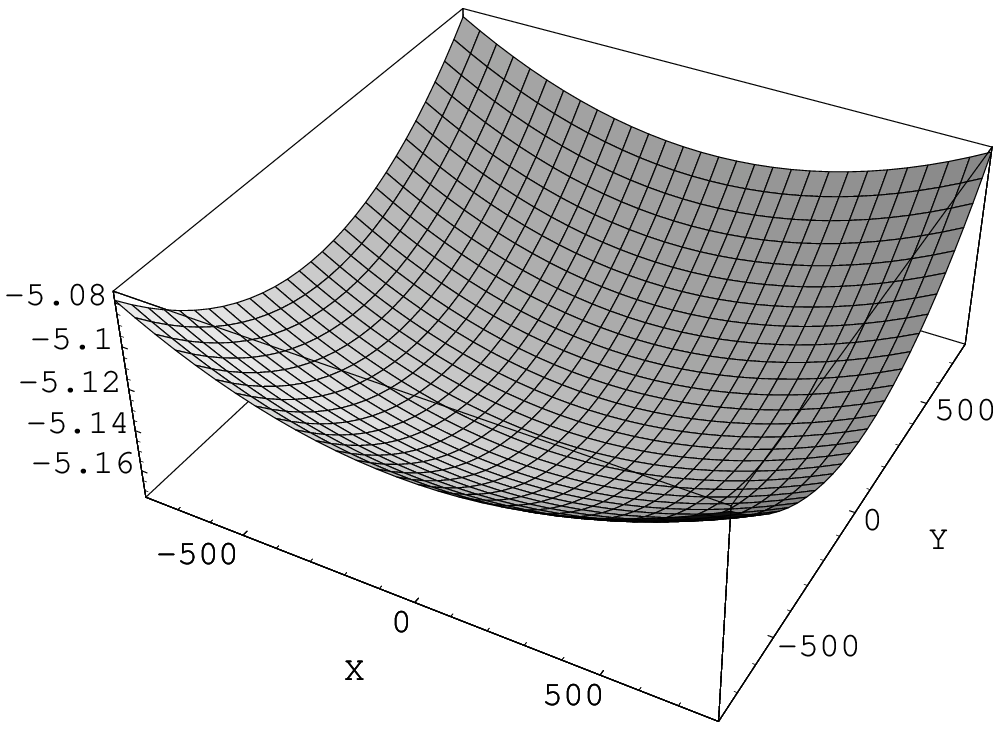}}\subfigure[]{\includegraphics[%
  width=7cm,
  keepaspectratio]{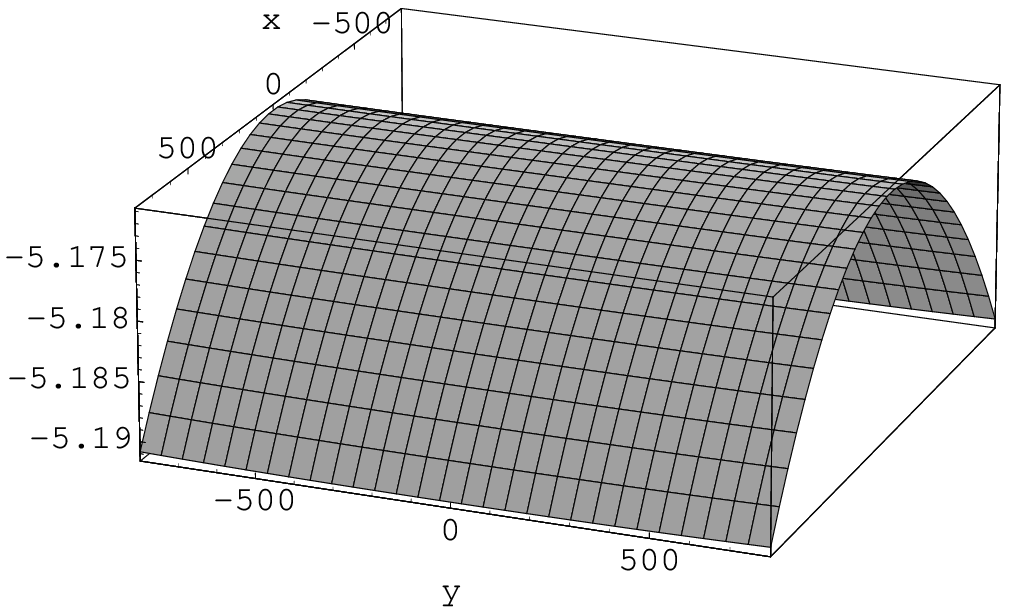}}\vspace{-0.5cm}\end{center}

\caption{Cuts through the generalized potential near the extremum,$\mathcal{V}(X,Y,0;\mathbf{0})$
(a) and $\mathcal{V}(\mathbf{0};x,y,0)$ (b), exemplified for $K=1.1K_{cr}$.
The $(XY)$ potential is almost harmonic. The $y$ direction virtually
coincides with the zero mode, whereas the potential exhibits an instability
with respect to $x$. \label{cap:V_xy(K1.1)}}
\end{figure}
\begin{figure}
\begin{center}\subfigure[Cut through the Z direction]{\includegraphics[%
  width=6cm,
  keepaspectratio]{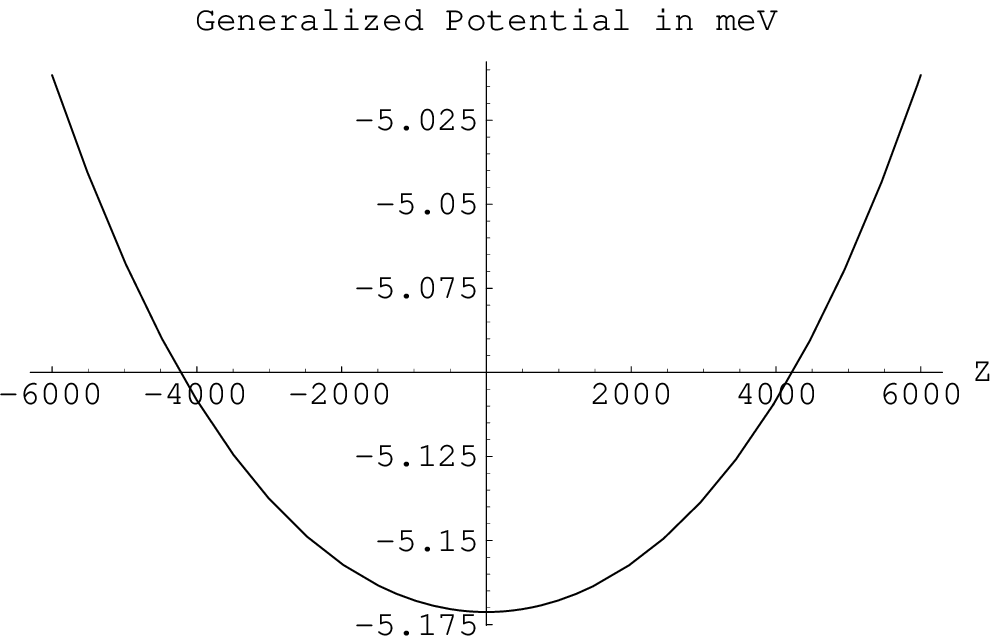}}\subfigure[Cut through the z direction]{\includegraphics[%
  width=6cm,
  keepaspectratio]{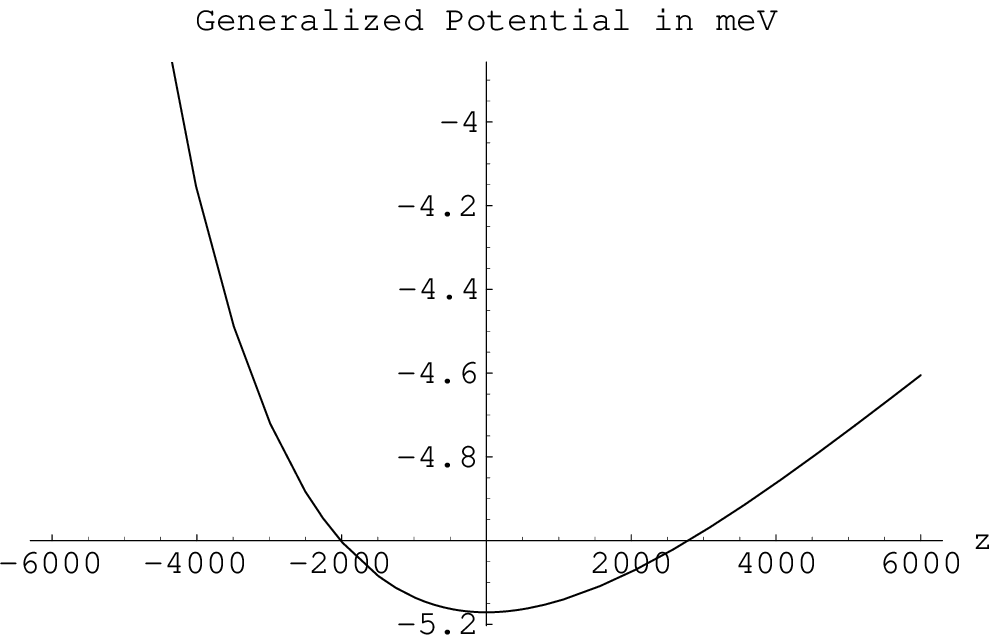}}\vspace{-0.5cm}\end{center}

\caption{Cuts through the generalized potential parallel to the magnetic field:
(a) $\mathcal{V}(0,0,Z;\mathbf{0})$ is roughly harmonic, (b) $\mathcal{V}(\mathbf{0};0,0,z)$
is strongly flattened on the right-hand region of the minimum. (Note
that the extremum is shifted to the origin.) \label{cap:Cuts-V(Z)}}
\end{figure}
 Even though the relative motion perpendicular to \textbf{B}, \textbf{}$\mathbf{s}_{\perp}\equiv(x,y)$,
is also gyrationally stabilized, there is no confining term for it
in the generalized potential as for the ECM. In fact, a look at the
generalized potential (Figure \ref{cap:V_xy(K1.1)}b) reveals that
$Q_{0}$ is a saddle point. The $y$ direction is the zero mode, while
the cut through the $x$ direction refers to a maximum. In this light,
we cannot expect the system to be arbitrarily stable in these two
degrees of freedom. This is a general fact, but for $K=1.1K_{cr}$
it is very pronounced. The reason being, near the extremum, the Coulomb
interaction (the source of these instabilities) becomes negligible
for higher $K$. A line of reasoning closely connected uses the time
scale at which the instability will have an impact. As the Coulomb
modes $\Omega(K)$ increase dramatically as $K\rightarrow K_{cr}^{+}$,
their characteristic time $T(K)=2\pi/\Omega(K)$ becomes much shorter
(cf. Table \ref{cap:scales}).%
\begin{table}
\begin{tabular}{|c|c|c|}
\hline 
$K/K_{cr}$&
$T_{\mathrm{C}}(K)$&
$|X_{0}(K)|$\tabularnewline
\hline
\hline 
1.1&
$.1-.4\,\mathrm{ns}$&
$1.4\cdot10^{4}\,\mathrm{a.u.}$\tabularnewline
\hline 
2&
$.4-1\,\mathrm{ns}$&
$2.9\cdot10^{4}\,\mathrm{a.u.}$\tabularnewline
\hline 
10&
$5-15\,\mathrm{ns}$&
$1.5\cdot10^{5}\,\mathrm{a.u.}$\tabularnewline
\hline
\end{tabular}

\caption{Summary of time scale of the Coulomb motion $(T_{\mathrm{C}})$ and
length scale $(|X_{0}|)$ for different values of pseudo-momentum
$K$\label{cap:scales}}
\end{table}

\begin{figure}
\subfigure[]{\includegraphics[%
  width=7.5cm,
  keepaspectratio]{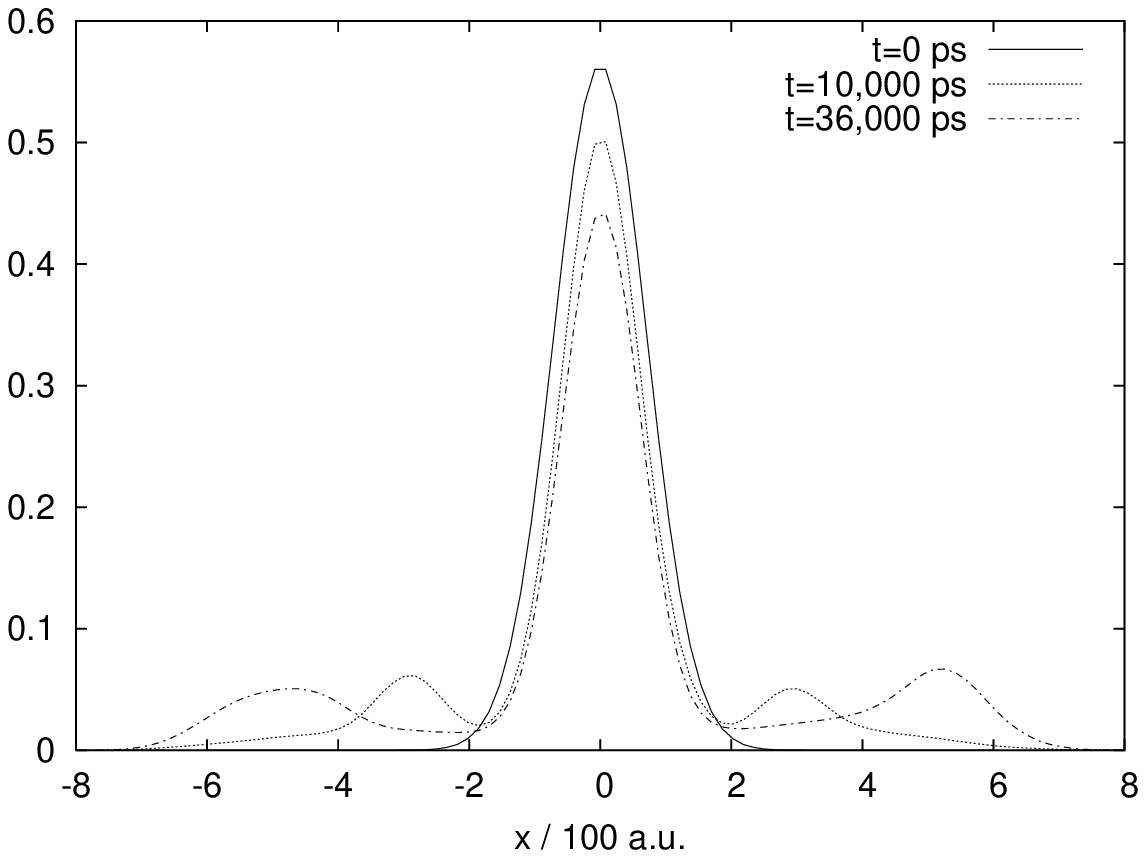}}\subfigure[]{\includegraphics[%
  width=7.5cm,
  keepaspectratio]{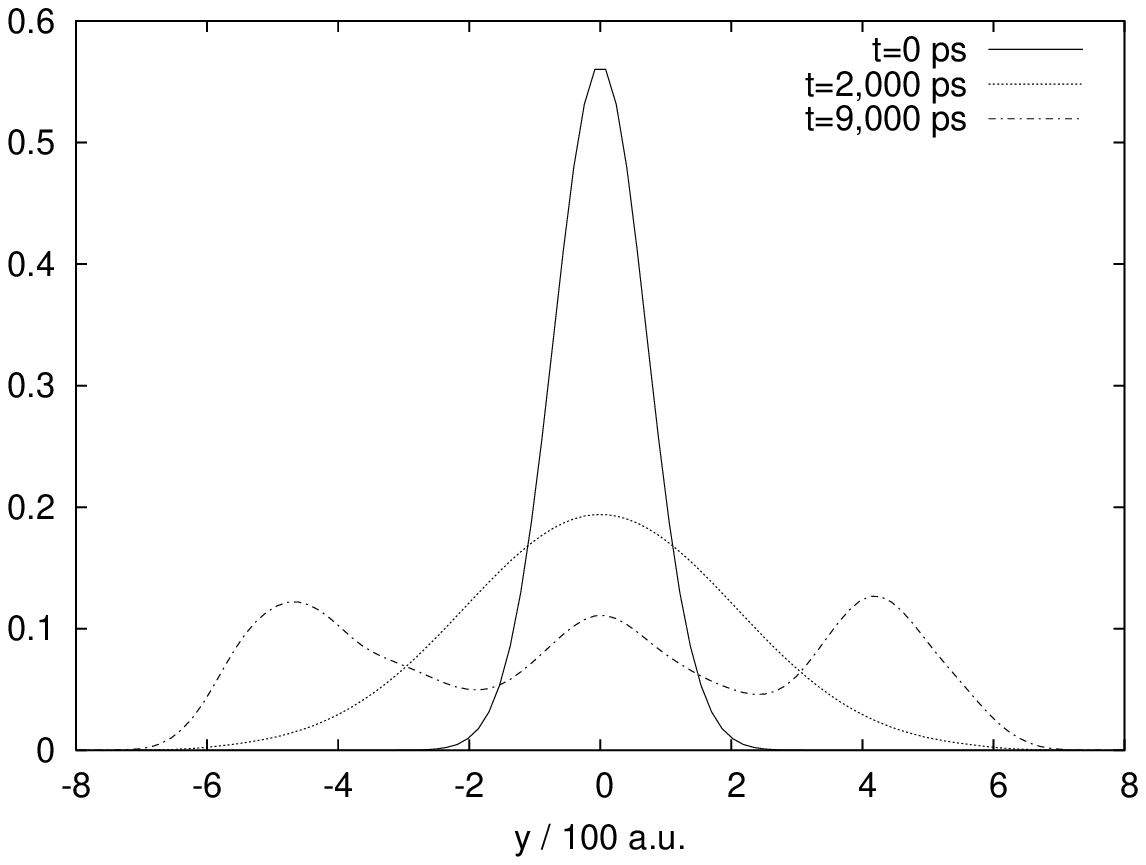}} \vspace{-0.5cm}

\subfigure[]{\includegraphics[%
  width=7.5cm,
  keepaspectratio]{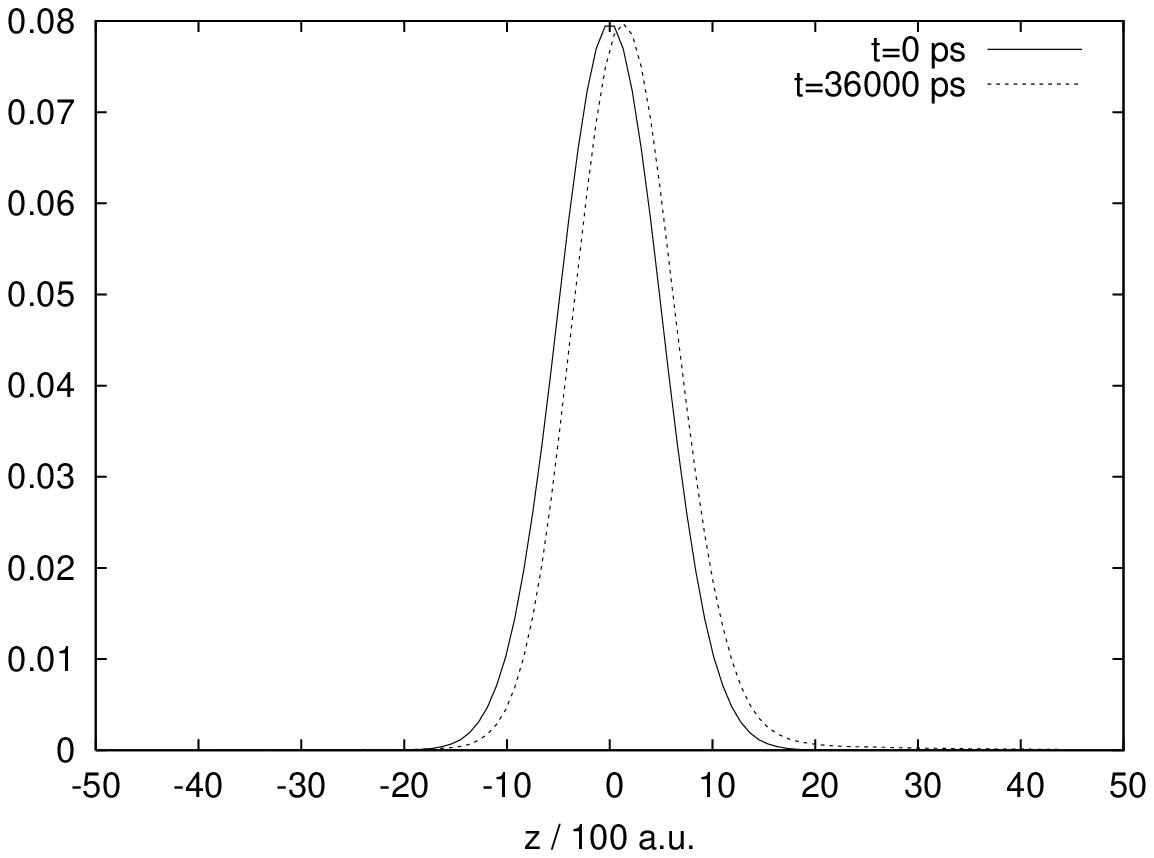}}\subfigure[]{\includegraphics[%
  width=8cm,
  keepaspectratio]{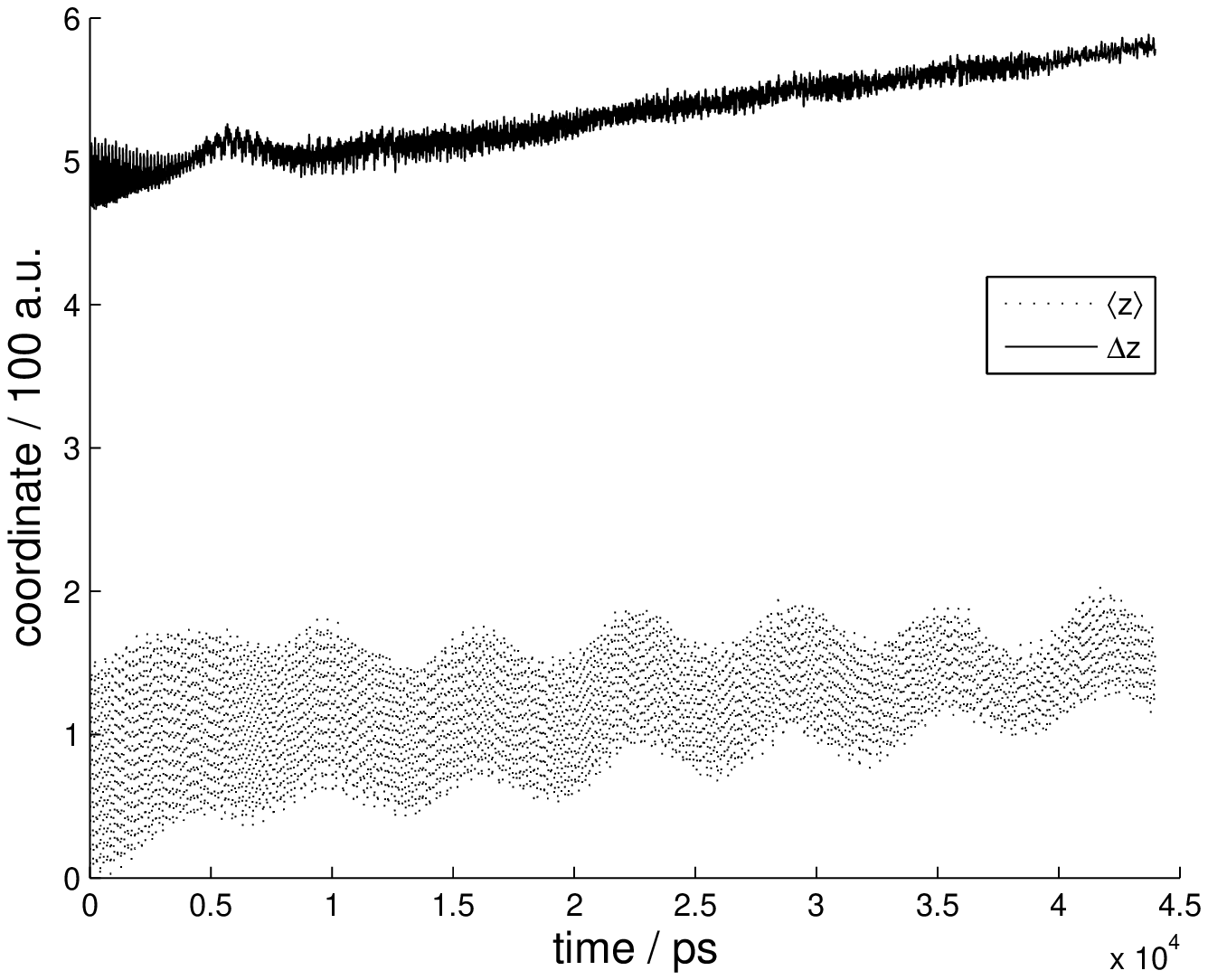}}

\caption{Snapshots of the wave packet's motion in the direction of $x$ (a),
$y$ (b) and $z$ (c) for $K=1.1K_{\mathrm{cr}}$. The plots display
the reduced densities $\rho_{x}$, $\rho_{y}$ and $\rho_{z}$ at
different times. While the unstable mode $x$ leaks slowly, the zero
mode $y$ spreads much more quickly. The wave packet in $z$ is only
marginally shifted over time. In (d), the different oscillations in
$z$ are illustrated.\label{cap:K1.1_xyz_snap}}
\end{figure}
These considerations shed light on Figure \ref{cap:K1.1_xyz_snap}.
For instance, $\rho_{x}$ soon develops two small side maxima. They
become more pronounced as time goes on and slowly spread towards the
boundary of the grid. The zero mode ($y$) spreads out much faster---that
is, to a significant extent after 2,000~ps already.

\begin{figure}
\begin{center}\includegraphics[%
  width=8cm]{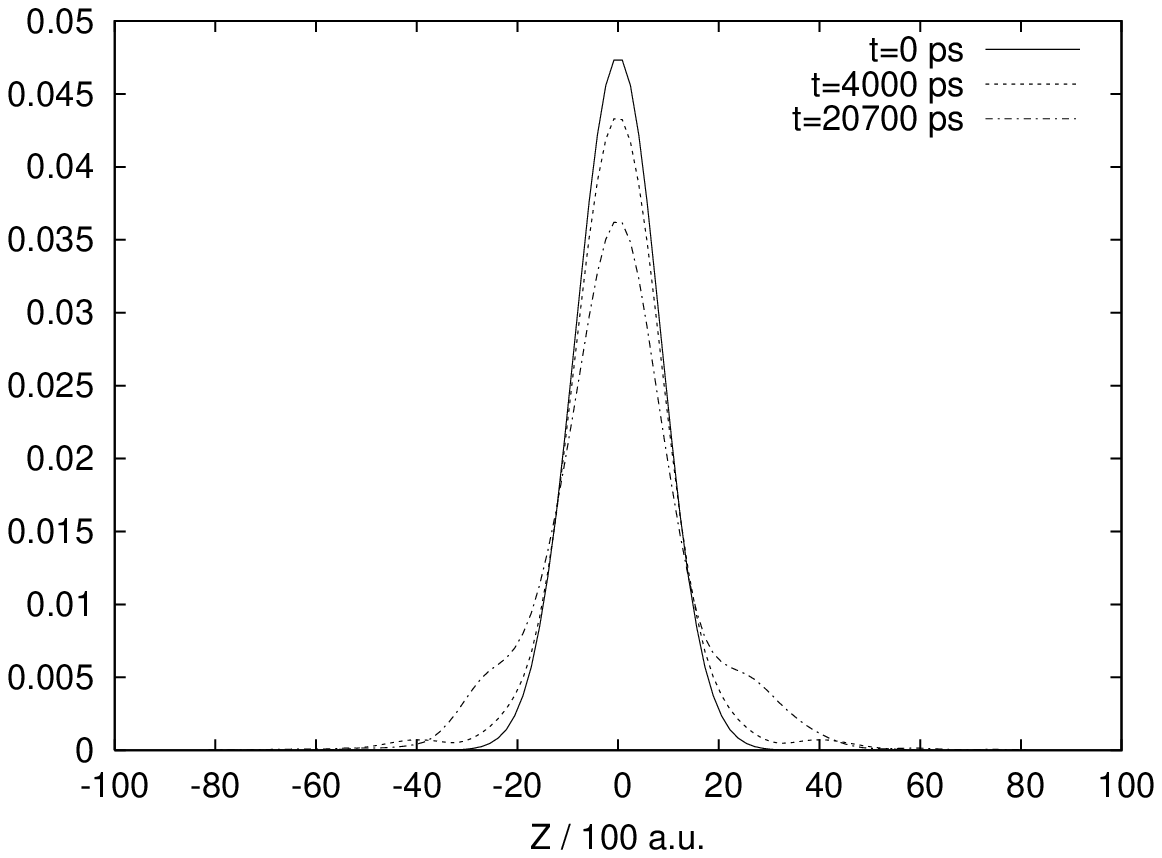}\end{center}

\caption{The motion of the $Z$ mode in the case $K/K_{cr}=1.1$. Three snapshots
of the reduced density $\rho_{Z}$ are plotted.\label{cap:K1.1Z_snap}}
\end{figure}
The reduced density for $Z$ (Fig. \ref{cap:K1.1Z_snap}) broadens
and continually leaks toward the grid ends. Although in the $Z$ direction,
the particle lives in a fairly harmonic well (Fig. \ref{cap:Cuts-V(Z)}),
it is not only lifted in energy to way above the bottom of the well,
but also affected by the instability in $x$ via coupling. Turning
to the last degree of freedom, $z$, we find an overall stable behavior.
Note that there is an overlay of a fast but tiny oscillation of the
packet's center and a slower and less regular one visible in Figure
\ref{cap:K1.1_xyz_snap}(d) as the envelope of $\langle z\rangle(t)$.

\paragraph{The case $K=2K_{cr}$:\label{par:K2.0}}

\begin{figure}
\vspace{-.5cm}

\hspace{-1.5cm}\subfigure[]{\includegraphics[%
  width=8cm,
  keepaspectratio]{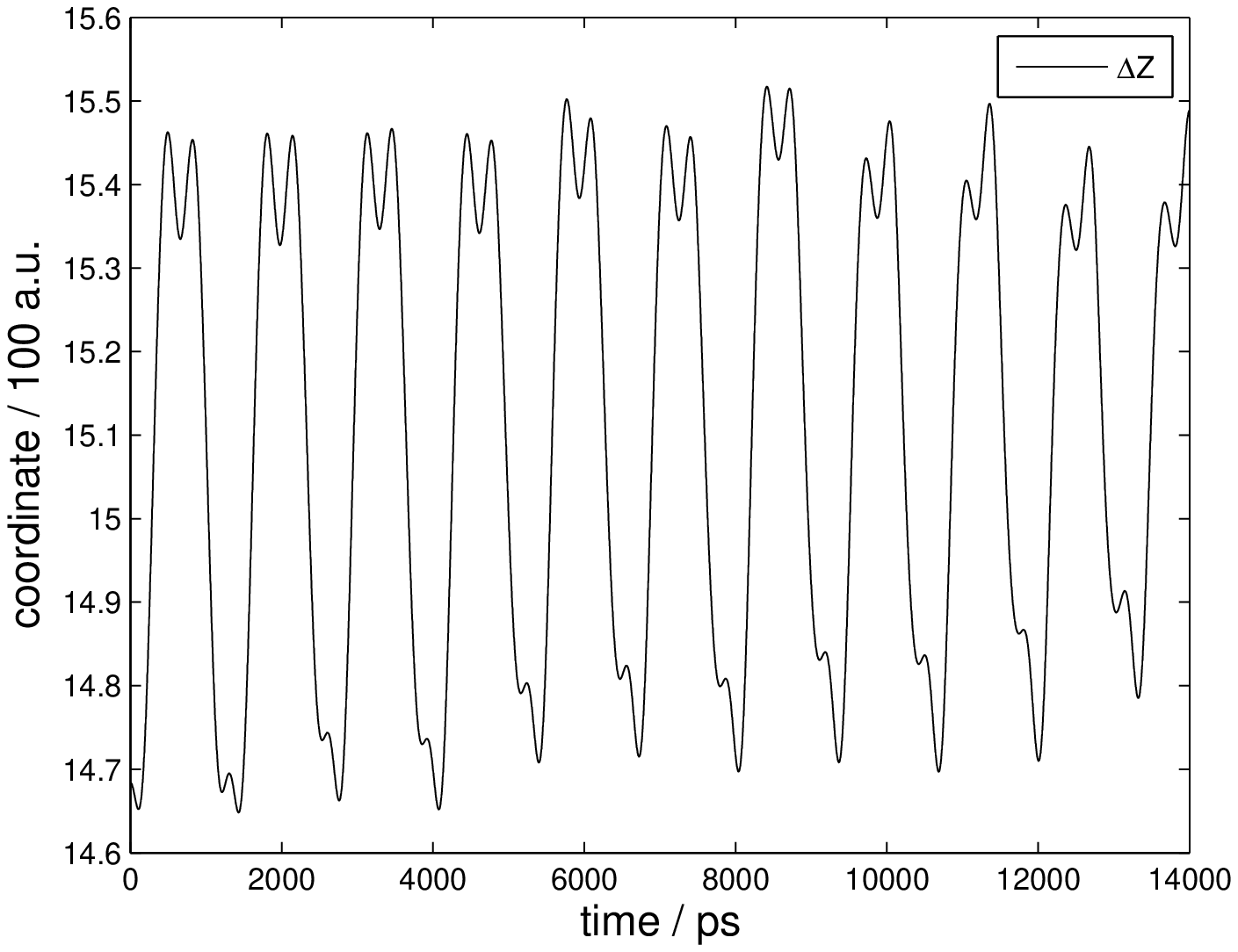}}\subfigure[]{\includegraphics[%
  width=8cm,
  keepaspectratio]{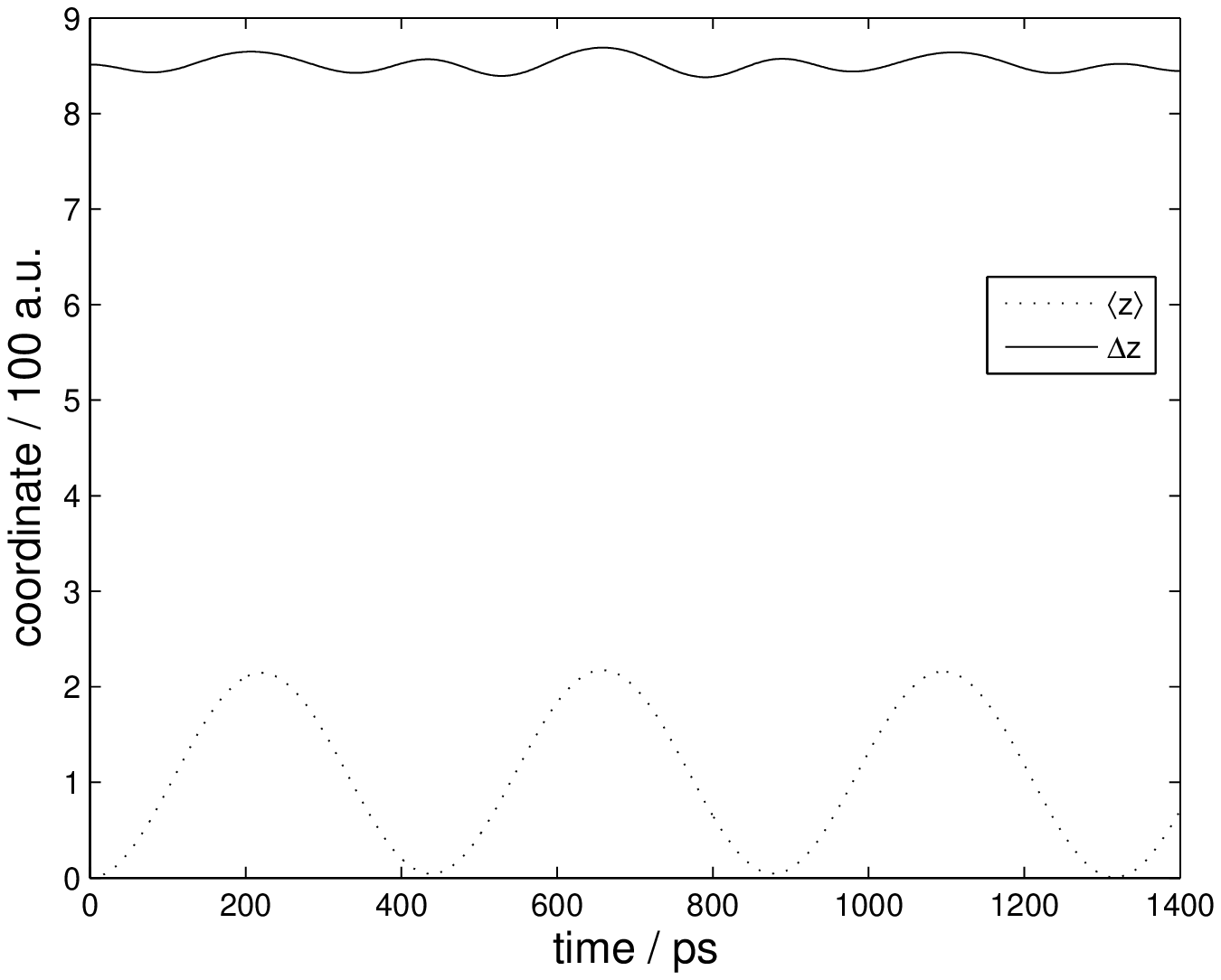}}

\caption{The expectation values and uncertainties over selected time periods
for the coordinates (a) $Z$ and (b) $z$ ($K/K_{cr}=2$).\label{cap:K2.0Zz}}
\end{figure}
For $K$ being twice its critical value, the vertical configuration
is virtually stable on a scale of $T\sim10^{4}\,\mathrm{ps}$. The
degrees of freedom $X$ and $Y$, as well as the parallel motion $Z$
and $z$, turn out to be indeed perfectly stable. Again, there are
two different types of tiny oscillations, displayed in Figure \ref{cap:K2.0Zz},
which are fingerprints of the coupling with the Coulombic motion (indicated
by plotting $\exv{q_{\kappa}}(t)$ over some periods $T_{Z}$, for
instance). %
\begin{figure}
\begin{center}\hspace{-0.5cm}\includegraphics[%
  width=7.5cm,
  keepaspectratio]{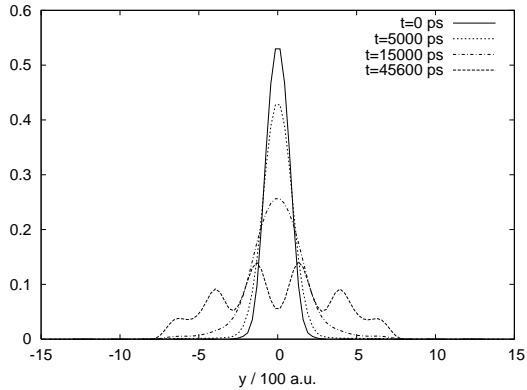}\vspace{-0.5cm}\end{center}

\caption{The relative coordinate $\mathbf{s}_{\perp}$ in the case $K/K_{cr}=2$:
Characteristic snapshots of the reduced density $\rho_{y}$.\label{cap:K2.0xy}}
\end{figure}

As opposed to the previous paragraph $K/K_{cr}=1.1$, the relative
motion in $x$ conveys a fairly stable if erratic impression. Still,
the wave packet in $y$ spreads as it did before (Fig. \ref{cap:K2.0xy}).
But keep in mind that it occurs at a much greater time scale than
in Fig. \ref{cap:K1.1_xyz_snap}(b): for example, after 5,000~ps
the broadening has only just begun to become significant.

\begin{figure}
\begin{center}\subfigure[]{\includegraphics[%
  width=7cm,
  keepaspectratio]{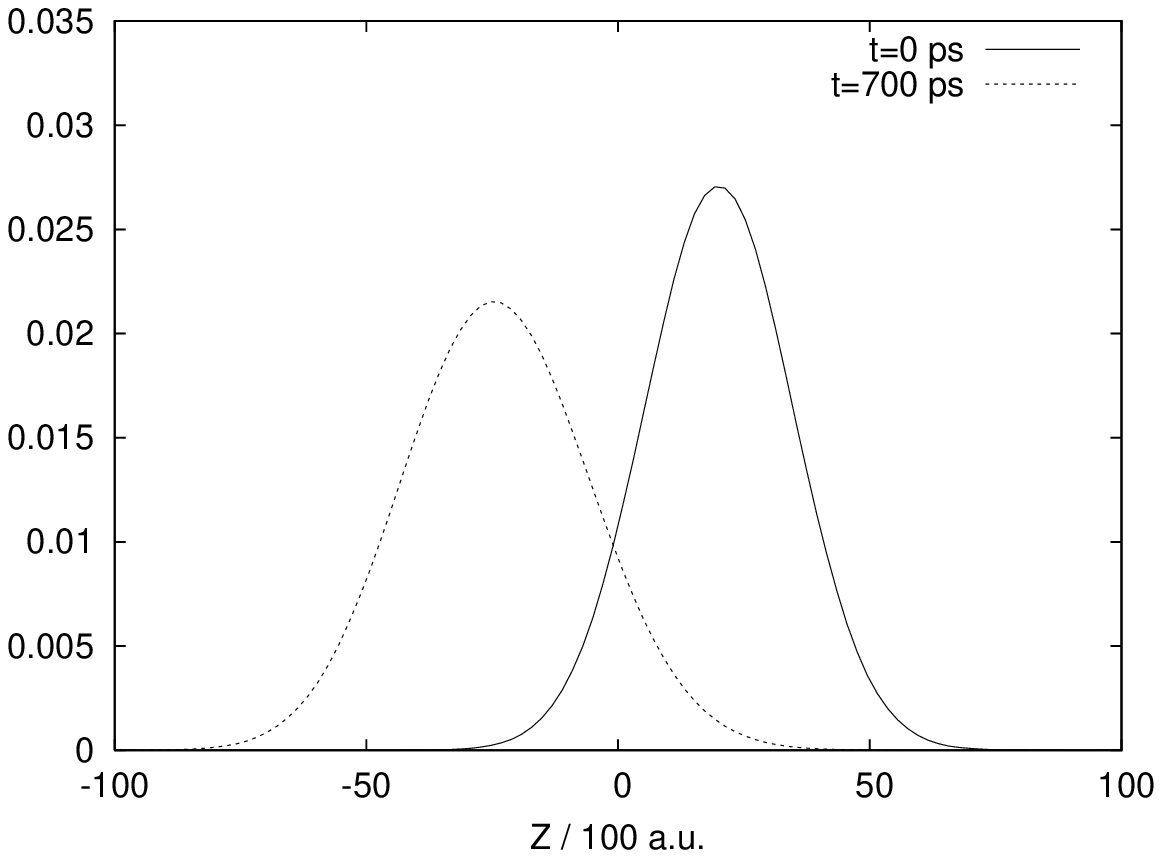}}\subfigure[]{\includegraphics[%
  width=7cm,
  keepaspectratio]{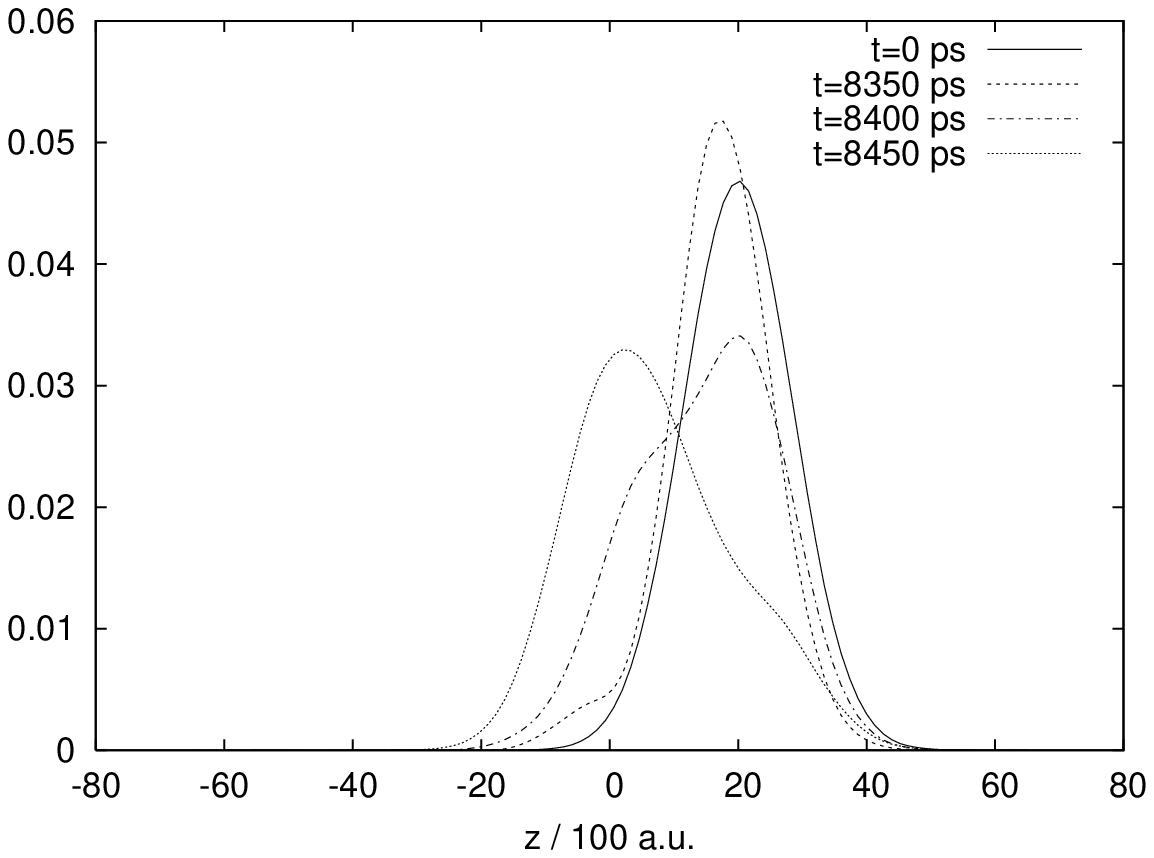}}\vspace{-1cm}\end{center}

\begin{center}\subfigure[]{\includegraphics[%
  width=7cm,
  keepaspectratio]{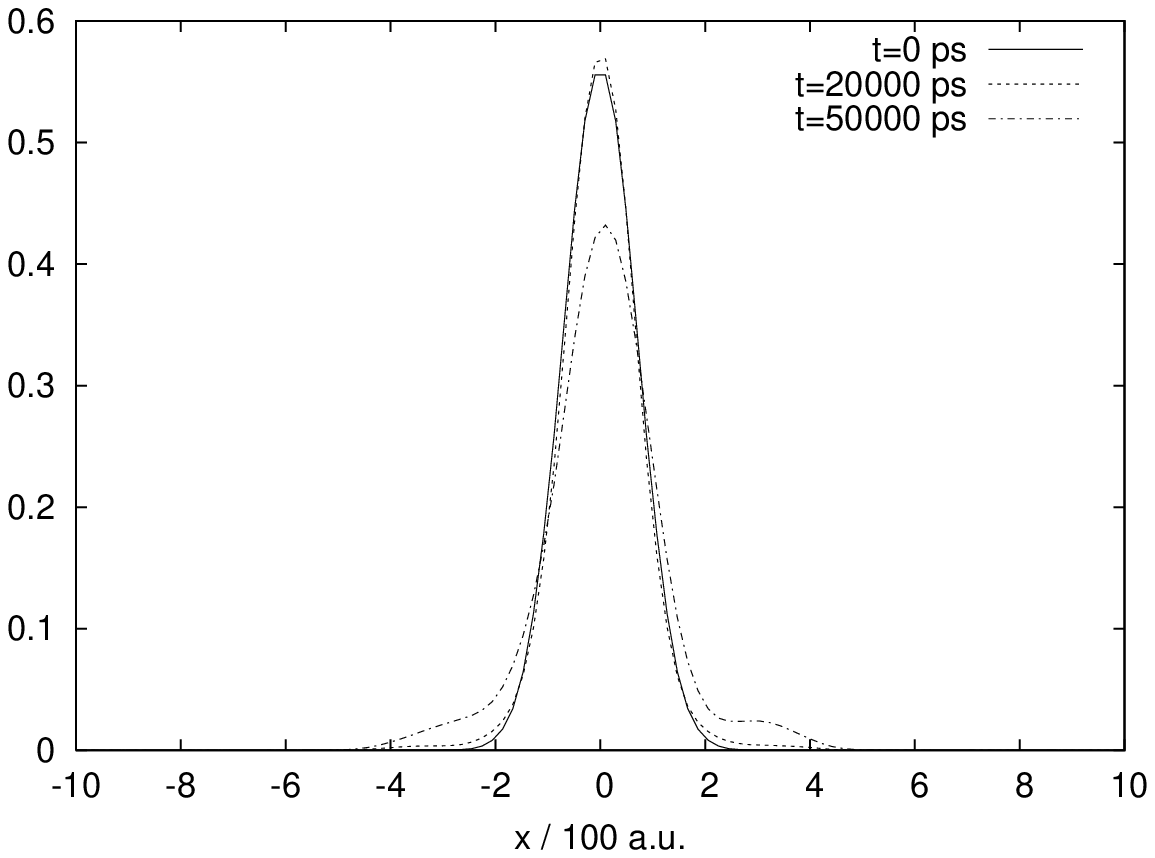}}\subfigure[]{\includegraphics[%
  width=7cm,
  keepaspectratio]{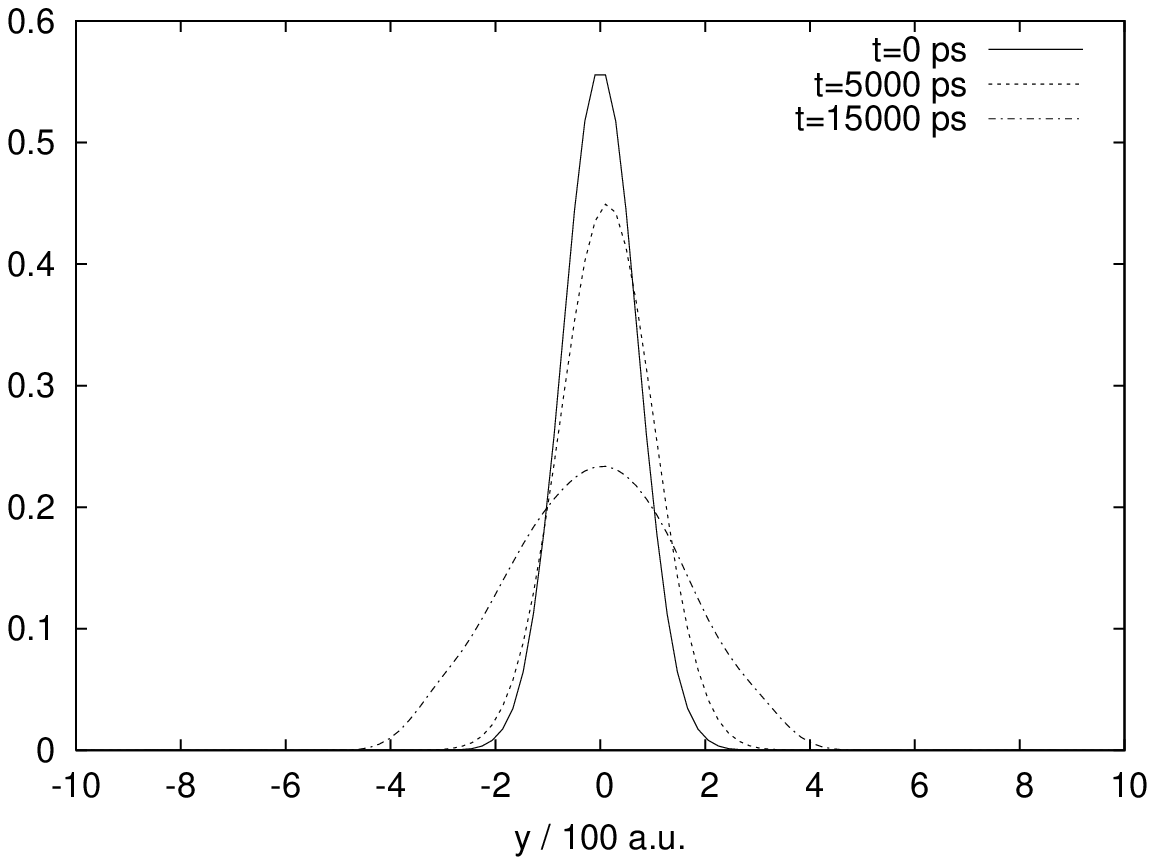}}\vspace{-0.5cm}\end{center}

\caption{Snapshots of the wave-packet upon an initial displacement by $Z=z=2,000\,\mathrm{a.u.}$
for $K/K_{cr}=2$. Plot (a) illustrates the strong deformations in
$Z$ in the course of the oscillation; (b) shows how the the $z$
packet is distorted after several thousand pico-seconds. The non-stable
modes $x,y$ are affected via coupling, see (c) and (d).\label{cap:K2.0_2Z}}
\end{figure}

A closer look at the response of the system upon \emph{displacing}
$Z$ and $z$ by $2,000\,\mathrm{a.u.}$ unveils that for the excited
degrees themselves, $Z$ and $z$, the wave packet is simply reflected
between two positions $\pm\exv{z}_{0}$ with minor ($Z$) or more
pronounced ($z$) deformations and smearing-out due to competing modes,
cf. Figure \ref{cap:K2.0_2Z}(a-b). The most interesting question
may be the effect on the $(xy)$ modes. In fact, $x$ is rendered
slightly unstable by the excitation of the parallel motion. The snapshots
taken at 20,000~ps and 50,000~ps, respectively, document how the
wave packet slowly but inevitably starts leaking, Fig. \ref{cap:K2.0_2Z}(c-d).
The lifetime of the $y$ mode is not drastically changed altogether.

\begin{figure}
\begin{center}\vspace{-0.5cm}\includegraphics[%
  width=8cm,
  keepaspectratio]{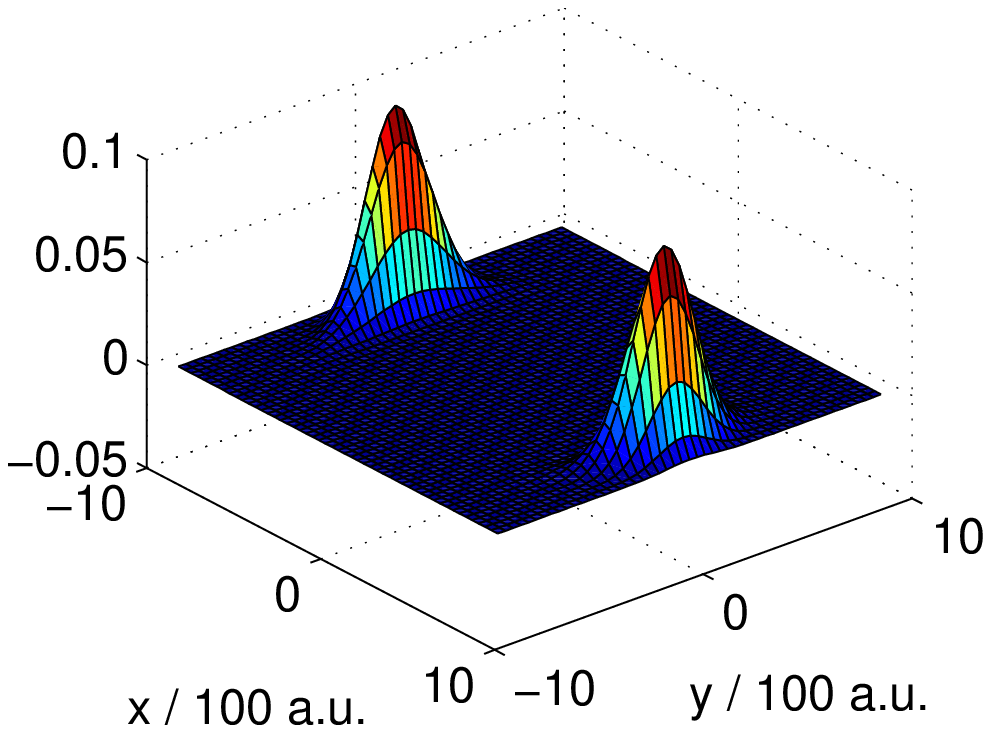}\vspace{-0.5cm}\end{center}

\caption{\label{cap:K2.0_relax}A generic relaxation state ($K=2K_{cr}$,
see text) -- plotted is the 2D-reduced density for the plane perpendicular
to $B$: the wave packet is driven away from the $x$ axis with increasing
basis size.}
\end{figure}
We also investigated the supposed ground state obtained via \emph{improved
relaxation}. It is understood that at best the exact system displays
a resonance state, so the quest for \emph{eigenstates} is meaningful
only insomuch as those of the discretized Hamiltonian may be interpreted
as localized states which really correspond to resonances. Our results
turned out to be very sensitive to both the primitive and the single-particle
basis sizes. Nonetheless, for a high if manageable accuracy, the pattern
that emerged was the following. Apart from some smaller deviations
from the initial state (Landau-orbital / harmonic-oscillator product),
a drastic alteration takes place in $\mathbf{s}_{\perp}\equiv(x,y)$
(Fig. \ref{cap:K2.0_relax}). The wave packet is split into two humps
that seem to be driven outwards until the basis size is exhausted,
while smeared out along the zero mode $y$.

This behavior raises the question how that squares with the potential
picture, given that the local instability characterized by $\omega_{x}^{2}\equiv w_{xx}/m$
is much weaker in magnitude than $\omega_{B}^{2}$. In order to clarify
these issues, we constructed a 2D model Hamiltonian via the natural
inclusion $H_{xy}:=H\!\mid_{\mathbf{R}=\mathbf{0};\, z=0}$. The study
of this toy model, which is numerically far more amenable, revealed
that it is capable of recovering many of the key features of the full
system. It therefore provided a valuable tool in detecting the key
to this mechanism---the paramagnetic term $\omega_{B}\ell_{z}$. To
begin with, the instability of the generalized potential in $x$ tempts
the packet to split up and be driven {}`downhill'. However, if there
were no paramagnetic term or if the same instability occurred isotropically,
this would have no discernible effect. However, owing to the anisotropy
of $\mathcal{V}$, $\boldsymbol{\ell}$ is no longer conserved:\[
\dot{\boldsymbol{\ell}}=-\mathbf{s}\times\nabla\mathcal{V}.\]
It is energetically favorable for the particle to go to ever lower
$\exv{\ell_{z}}<0$ while being expelled from the unstable line $x=0$,
thus introducing a high correlation between $x$ and $y$.

\begin{figure}
\begin{center}\vspace{-0.5cm}\includegraphics[%
  width=8cm,
  keepaspectratio]{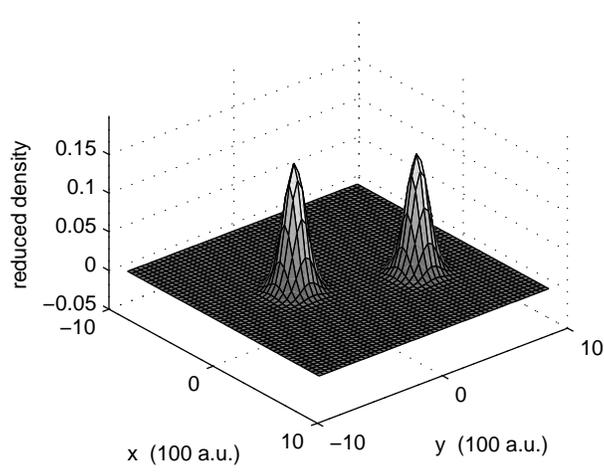}\vspace{-0.5cm}\end{center}

\caption{\label{cap:K2.0_2D}Wave packet of the 2D model ($xy$) after $T=160\,\mathrm{ns}$
($K=2K_{cr}$, see text)}
\end{figure}

Note that this effect essentially amounts to the fact that the giant-dipole
resonance state is far from the energetically lowest configuration.
That does not pose an immediate constraint on the system's lifetime
under \emph{propagation}, since our initial Landau/harmonic-oscillator
state has a virtually zero overlap with the above {}`relaxation state'.
Hence in a propagation, this {}`bimodal' relaxation state is not
accessible, which explains why the system is so much more stable than
its relaxation suggests. Physically speaking, the cyclotron gyration
field stabilizes the motion near the stationary configuration for
\emph{fixed} energy and thus inhibits a delocalized state at a \emph{lower}
energy. -- This is approved by a 2D numerically exact time evolution
(Fig. \ref{cap:K2.0_2D}), which reveals no breakup for many nanoseconds,
until the wave packet is slightly rotated in the $xy$ plane and split
up roughly in the $y$ direction---and not in $x$ as in the relaxation
case! Note that it acquires a small but non-zero value of $\exv{\ell_{z}}\sim-1$,
in contrast to $\exv{\ell_{z}}\sim-60$ in Fig.~\ref{cap:K2.0_relax};
this also reflects the higher level of stability. It is only after
way more than $100\,\mathrm{ns}$ that the packet hits the grid's
boundary. These outcomes seem to confirm that for moderate values
of $K$, our system is quasi-stable on a time scale of up to $T\sim100\,\mathrm{ns}$.%

\paragraph{The case $K=10K_{cr}$:}

For this value, which corresponds to $E\equiv BK/M=207\,\mathrm{\frac{kV}{m}}$,
we find the system to be practically stable on the time scale we considered,
$T=100\,\mathrm{ns}$. 

\begin{figure}
\begin{center}\subfigure[]{\includegraphics[%
  width=7.5cm,
  keepaspectratio]{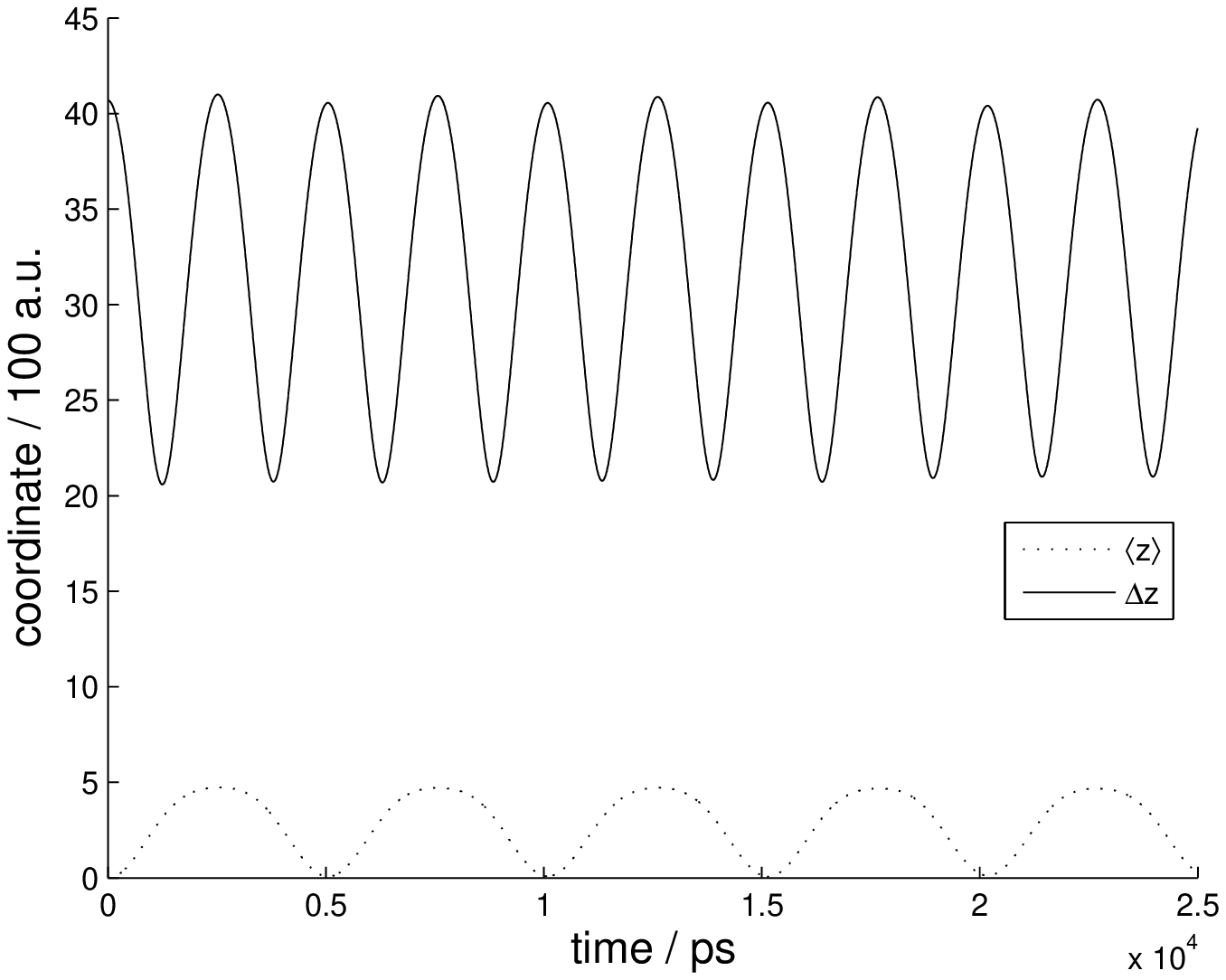}}\hspace{-0.5cm}\subfigure[]{\includegraphics[%
  width=7.5cm,
  keepaspectratio]{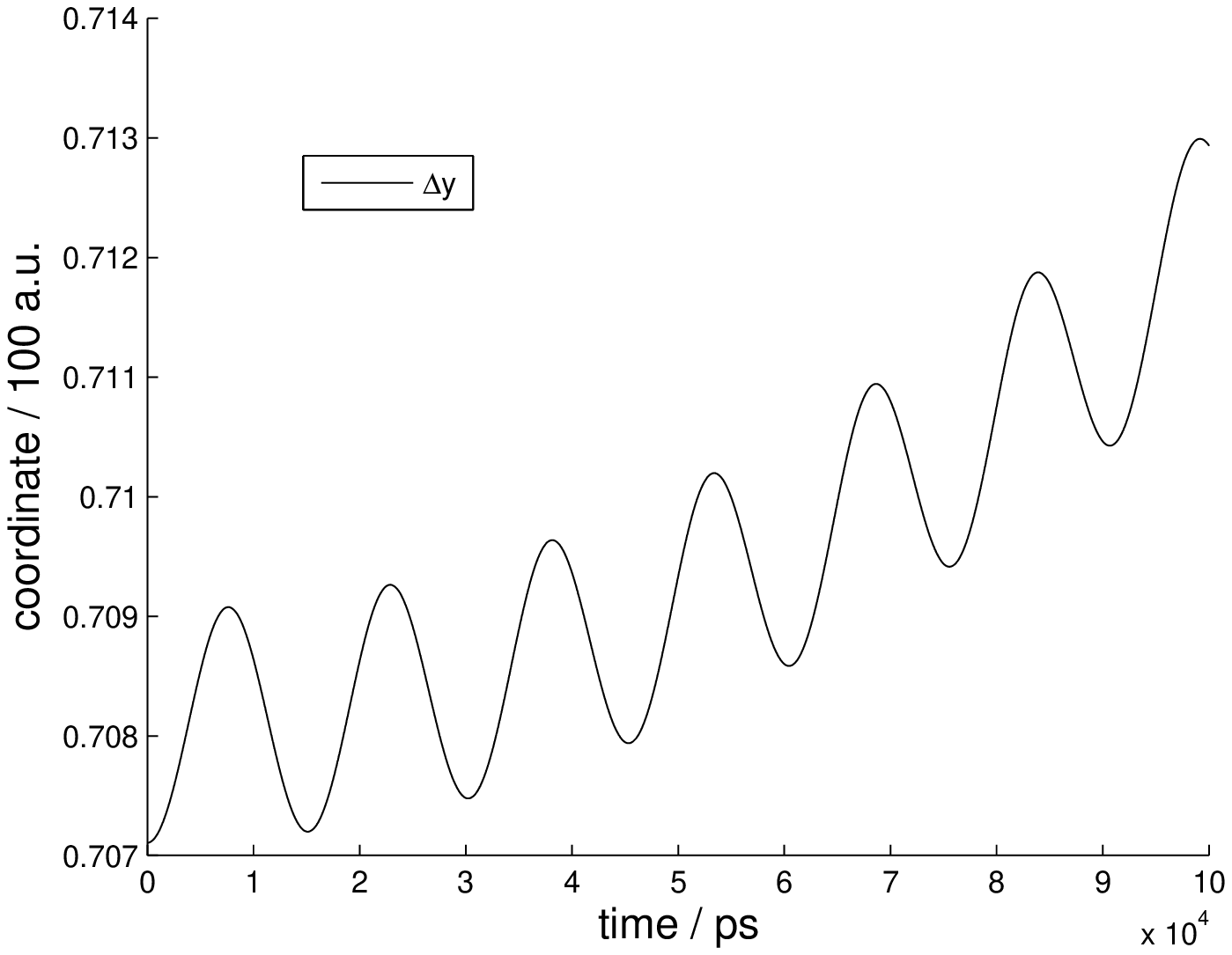}}\vspace{-0.5cm}\end{center}

\caption{The time evolution of $\exv{z},\Delta z$ (a) and $\Delta y$ (b)
for $K/K_{cr}=10$. Both degrees are altogether stable, but oscillate
due to coupling. A marginal broadening can be seen in $y$. \label{cap:K10.0zy}}
\end{figure}
The initial wave packet experiences only tiny deformations in most
degrees of freedom, which are not discussed here. However, there are
oscillations for $z$, which are more pronounced but still marginal
compared to the spatial extension of the decentered state; these are
sketched in Figure \ref{cap:K10.0zy}(a). While its center shows a
shift to positive values of $z$, its elongation is on the order of
$250\,\mathrm{a.u.}$, and the width varies by a factor of two, all
at a period of about $T_{z}$. In contrast with $x$ and any other
degree, the width in the $y$ direction shows a steady if minute broadening
(Fig. \ref{cap:K10.0zy}b). This might be taken for a sign that eventually,
the system is bound to decay. Then again, this takes $300,000\,\mathrm{ps}$
to move by one atomic unit and is beyond the time scale regarded here.

After having come a long way to find the system practically stable
for high enough values of the pseudo-momentum, we now explore the
robustness with respect to displacements. Skipping the detailed behavior,
we find that the overall stability was not affected. Due to coupling
to other modes, the excitation of the parallel motion imprints some
oscillations both of the packet's center and its width on the perpendicular
degrees of freedom.

Lastly, applying a relaxation revealed similar phenomena as in the
previous paragraph. Regardless of the specific shape of the relaxation
state, it can be reasoned that the instabilities of course still underlie
the system, but that the lifetimes for sufficiently large $K$ are
too large to be observed in our propagation.

\subsubsection{Influence of the rotational freedom}

To complete the discussion, let us now investigate the giant dipole
states for different settings of the extremal angle $\phi_{0}$ of
the relative coordinate $\mathbf{s}_{0}$ on the circle. So far, the
focus was on $\phi_{0}=\pi/2$ only (or, equivalently, $\Delta=-\frac{\pi}{2}$),
which we referred to as the {}`vertical configuration'. In the normal-mode
analysis, however, we found an interesting frequency pattern as we
varied this extremal angle. It was argued that each $\phi_{0}$ corresponds
to a different extremal point, i.e., for each value of $(B,K)$ there
is really a one-parameter class of extremal points $Q(\phi_{0})$
with different stability and spectral properties.

\begin{figure}
\begin{center}\includegraphics[%
  width=8cm,
  keepaspectratio]{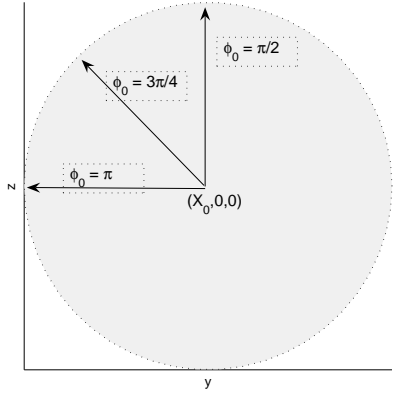}\vspace{-0.5cm}\end{center}

\caption{The circular configuration $\mathbf{s}^{(0)}=s_{0}(0,\cos\phi_{0},\sin\phi_{0})^{T}$
for $N=2$, depending on the angle $\phi_{0}$. The arrows refer to
$\mathbf{s}^{(0)}$ in the $yz$-plane for different values of $\phi_{0}$.
The $X$ axis points into the page.\label{cap:circle}}
\end{figure}
As examples, we thus investigate both the supposedly unstable {}`horizontal
configuration' $\phi_{0}=\pi$ $(\Delta=0)$ and a {}`diagonal configuration'
$\phi_{0}=\frac{3}{4}\pi$ $(\Delta=-\pi/4)$, as visualized in Figure
\ref{cap:circle}. This was done for the special cases of $K/K_{cr}=2,\,10.$

\begin{figure}
\begin{center}\includegraphics[%
  width=7.5cm]{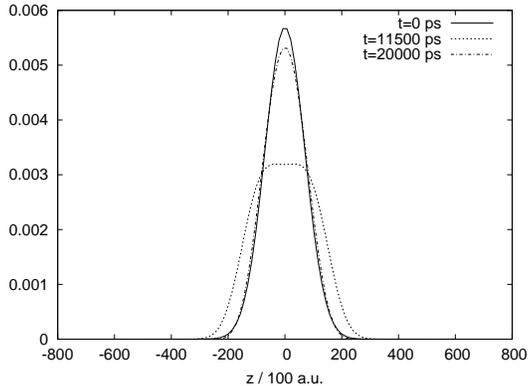}\vspace{-0.5cm}\end{center}

\caption{The horizontal configuration $\phi_{0}=\pi\,(K/K_{cr}=10)$. The
wave packet switches between two positions (snapshots of reduced density
for 11,500 ps and 20,000 ps).\label{cap:K10.0_phi1.0}}
\end{figure}

To sum up our findings, the horizontal configuration indeed adds an
instability, which is discernible even for very high $K$, if less
distinct. The diagonal configuration was partly unstable on a timescale
comparable to that of the vertical case, $T\sim10^{4}\,\mathrm{ps}$.
A rotation of the extremum obviously rotates the zero mode, too, which
is why the detailed dynamics is a different one for the relative degrees
of freedom. For instance, in case that $\phi_{0}=\pi$, the zero mode
points along the $z$ direction, which would otherwise live in a harmonic
potential. As a consequence, even for $K/K_{cr}=10$ the $z$-packet
(prepared with a non-zero frequency $\omega_{z}=10^{-8}\mathrm{a.u.}$)
smoothly oscillates between its initial and a much more smeared-out
state according to Fig. \ref{cap:K10.0_phi1.0}.

\subsubsection{Spectrum}

\begin{figure}
\begin{center}\includegraphics[%
  width=8cm]{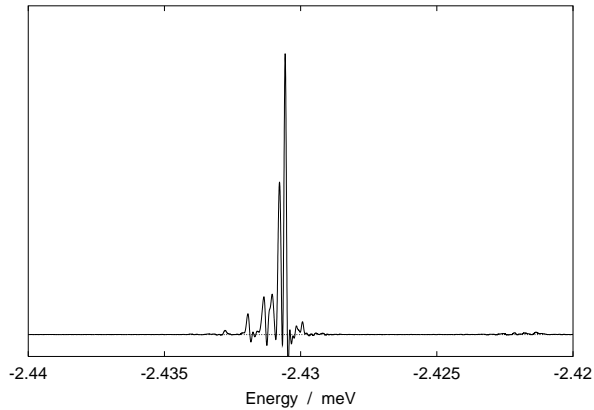}\vspace{-0.5cm}\end{center}

\caption{Fourier transform of the auto-correlation function (the {}`spectrum')
for $K/K_{cr}=1.1$. \label{cap:K1.1_spec}}
\end{figure}
We finally look into some spectral properties of the GDS for the different
values of $K/K_{cr}$. The spectrum calculated from a propagation
for $K/K_{cr}=1.1$ (Fig. \ref{cap:K1.1_spec}) looks somewhat fuzzy,
but the main peak is still distinct enough to vaguely resemble a resonance.
The situation is less ambiguous in the case $K/K_{cr}=2$ , as the
{}`decay' occurs on a much greater time scale.%
\begin{figure}
\begin{center}\includegraphics[%
  width=7.5cm,
  keepaspectratio]{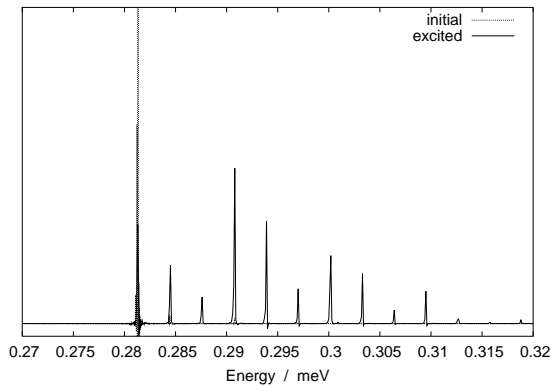}\vspace{-0.5cm}\end{center}

\caption{Spectrum for $K/K_{cr}=2$. The dotted peak again belongs to the
initial state. For the excited (or displaced) initial state, the spectrum
shows a sequence of peaks (here almost equidistant).\label{cap:K2.0_spec}}
\end{figure}
The spectrum in Fig.~\ref{cap:K2.0_spec} illustrates the resonance
character of the system. Our initial wave packet gives a peak at $E\simeq0.281\textrm{ meV}$,
whereas  the {}`displaced state' with $\exv{z}_{0}=\exv{Z}_{0}=2,000\,\mathrm{a.u.}$
produces a rather interesting excitation spectrum (same figure), revealing
overlaps with many eigenstates. The equidistant spacing of the peaks
might be interpreted as a signature of harmonicity in both excitations
($Z,\, z$). Finally, in the most stable case $K/K_{cr}=10$, the
spectrum consists of a sharp dominant peak at about 2.2422~eV, plus
a very tiny one 0.0017~eV above. This somehow quantifies our empirical
observation that $K/K_{cr}=10$ is quasi-stable.

\section{Conclusion and outlook}

We have studied the internal motion of $N$-electron atoms in crossed
magnetic and electric fields utilizing the \emph{generalized potential}.
For electric fields above a critical value, a \emph{strongly decentered}
potential well forms. It supports \emph{giant dipole states} where
the electrons' center of mass is aligned along the electric field,
with all electrons in a symmetric circular configuration in the orthogonal
plane. These states have been investigated for an arbitrary number
of electrons in a normal-mode analysis, and numerically for the case
of two-electron giant dipole states employing wave-packet propagation.

The normal-mode analysis of the $N$-electron atoms leads to a \emph{quadratic
eigenvalue problem}. Its eigenmodes (as well as the eigenvectors)
have been studied depending on the field strengths $B,\, E$ as well
as the electron number $N$. A subsequent classification according
to their characteristics indicated that the $3N$ modes were grouped
into $N$ \emph{cyclotron modes} and $N$ \emph{Coulomb modes}, corresponding
mostly to the motion induced by the magnetic field and the Coulomb
potential, respectively. Moreover, we found a \emph{CM mode} reflecting
the center-of-mass gyration in the magnetic field, as well as a \emph{zero
mode} referring to a rotational invariance of the extremal configuration,
also studied in detail. The residual modes were termed \emph{decay
modes} on account of their peculiar behavior suggesting instability
for atoms with three or more electrons. Their absence for the system
$N=2$ provided a strong indication of local stability, with the single
exception of the so-called \emph{horizontal configuration}.

In the second part of the present work, an ab-initio simulation of
the six-dimensional two-electron system was performed. Since this
type of investigation can be viewed as the state of the art of what
is numerically feasible, we resorted to the Multi-Configuration Time-Dependent
Hartree (\noun{mctdh}) method, a wave-packet dynamics tool known
for its unique efficiency in higher dimensions. Both the stability
of the six degrees of freedom and some spectral properties have been
examined. To establish the link to the normal-mode analysis, the influence
of the rotational freedom of the circular configuration has also been
investigated. We find that the stability of the system strongly increases
for larger electric fields. For a field strength twice the critical
value, some modes experience a decay on the time scale of $10^{4}-10^{5}\,\mathrm{ps}$;
for the tenfold critical value, there is no instablity on the scale
we considered, i.e., $10^{5}\,\mathrm{ps}$. These states proved rather
robust against perturbations, simulated by displacements parallel
to the magnetic field, where the motion is not gyrationally stabilized.
As anticipated, the stability of the giant dipole states turned out
to apply only outside the singular {}`horizontal configuration'
of the extremum.\\

While the investigation into the \emph{local} aspects of stability
can be regarded as somewhat completed, a rigorous numerical analysis
is to date limited with an eye toward time and computational effort.
First and foremost, the \noun{mctdh} calculations set up so far
could in principle be modified to the case of $N\geq3$ electrons.
However, this appears to be a massive numerical challenge, since the
\noun{mctdh} method is slowed down significantly the more degrees
of freedom are involved. One approach to circumvent this would be
an \emph{adiabatic separation} similar to \cite{bezchastnov02}, amounting
to an average over the rapid cyclotron motion. This would allow for
a significantly faster integration. For future investigations, it
may also be beneficial for short-time propagations to implement a
\emph{discrete-variable representation} of the Landau orbitals in
\noun{mctdh}. 

Apart from these questions, concerned with a wider range of application
and efficiency, the task arises to make the analysis more quantitative.
A systematic resonance calculation yielding both the energy levels
\emph{and} decay widths of the giant-dipole states is desirable, but
has so far proven to be a severe challenge computationally due to
the type of instability. It might also yield a deeper understanding
of the resonance wave functions. Ultimately, one could even go so
far as to simulate the \emph{preparation} of giant dipole states,
migrating from the Coulomb well to the decentered extremum. In analogy
to the scheme suggested for single electrons \cite{Ave99:3695}, this
may serve as a bridge to the experimental verification of giant dipole
resonances.

\bibliographystyle{prsty}
\cleardoublepage\addcontentsline{toc}{chapter}{\bibname}\bibliography{gd}

\end{document}